\documentclass[aps,prm,reprint,amsmath,amssymb]{revtex4-2}
\usepackage[utf8]{inputenc}
\usepackage[T1]{fontenc}
\usepackage[paper=a4paper,left=10mm,right=10mm,top=20mm,bottom=20mm]{geometry}

\usepackage{xcolor}
\usepackage{textgreek}
\usepackage[greek, english]{babel}
\usepackage{graphicx}
\usepackage{subcaption}
\usepackage{varwidth}
\usepackage{float}
\usepackage{siunitx}
\usepackage{braket}
\usepackage{url}
\usepackage{amsmath}
\usepackage{chngcntr}
\usepackage[format=plain, labelfont=bf, indention=0cm, aboveskip=20px, belowskip=0px]{caption}

\usepackage{hyperref}

\hypersetup{
	colorlinks   = true, % Colours links instead of boxes
	urlcolor     = blue, % Colour for external hyperlinks
	linkcolor    = blue, % Colour of internal links
	citecolor   = blue % Colour of citations
}

\newcommand{\chr}{black} % color for changes due to reviewer comments
\newcommand{\cho}{black} % color for changes on our own
\newcommand{\chm}{black} % color for martin
\newcommand{\tc}[2]{\textcolor{#1}{#2}}
\newcommand{\sub}[1]{\textsubscript{#1}}
\newcommand{\super}[1]{\textsuperscript{#1}}

\newcommand{\sfs}{0.49\linewidth} % sub-figure-scale
\graphicspath{{"C:/Users/AMP Jona/Documents/GaNius/graphics/ScN/"}}

\begin{document}
	
	\title{Band gaps and phonons of quasi-bulk rocksalt ScN}
	\author{Jona Grümbel}\email{jona.gruembel@ovgu.de}  
	\affiliation{Institut für Physik, Otto-von-Guericke-Universität Magdeburg, Universitätsplatz 2, 39106, Magdeburg, Germany}
	\author{Yuichi Oshima}\affiliation{Research Center for Electronic and Optical Materials, National Institute for Materials Science , 1-1 Namiki, Tsukuba, Ibaraki 305-0044, Japan}
	\author{Adam Dubroka} \affiliation{Department of Condensed Matter Physics, Masaryk University, Kotl\'a\v{r}sk\'a 2, 611-37 Brno, Czech Republic}
	\author{Manfred Ramsteiner} \affiliation{ Paul-Drude-Institut für Festkörperelektronik (PDI), Hausvogteiplatz 5-7, 10117, Berlin, Germany}
	\author{Rüdiger Goldhahn}\affiliation{Institut für Physik, Otto-von-Guericke-Universität Magdeburg, Universitätsplatz 2, 39106, Magdeburg, Germany}
	\author{Martin Feneberg}\affiliation{Institut für Physik, Otto-von-Guericke-Universität Magdeburg, Universitätsplatz 2, 39106, Magdeburg, Germany}
	
	\begin{abstract}
		ScN is an emerging transition metal nitride with unique physical properties arising from the $d$-electrons of Sc. In this work we present the results of optical characterization techniques spectroscopic ellipsometry, Raman spectroscopy, and photoluminescence measurements of a $\SI{40}{\micro\m}$ thick, fully relaxed, and only weakly n-type doped ($n=\SI{1.2E18}{\per\cm\cubed}$) ScN film deposited by halide vapor phase epitaxy (HVPE) on r-sapphire substrate. Spectroscopic ellipsometry yields an indirect bandgap of $\SI{1.1}{\eV}$ while the lowest direct interband transition is observed at $E_\mathrm{g,opt}=\SI{2.16}{\eV}$ in the dielectric function. A broad luminescence feature at $\SI{2.15}{\eV}$ is observed, matching this transition. We derive an estimate for the exciton binding energy ($E_{\mathrm{bX}}\approx\SI{14}{\meV}$) as well as the Born effective charges $Z^*_{\mathrm{Sc}}=-Z^*_\mathrm{N} = 3.78$. In the infrared spectral region we observe a strong phonon and a weak plasmon absorption. We precisely determine the transverse optical phonon eigenfrequency ($\omega_{\text{TO}}=\SI{340.7}{\per\cm}$), the high frequency dielectric constant ($\varepsilon_{\infty}=8.3$) and the static dielectric constant ($\varepsilon_{\text{stat}}=29.5$). \tc{\cho}{Raman measurements using various excitation energies show resonant multi-phonon scattering up to 6LO ($6th$ order overtone for longitudinal optical (LO) phonons) for excitation above the optical band gap ($E_{\text{Laser}}>E_{\mathrm{g},\text{opt}}$), where the allowed $2$LO scattering is the dominant scattering mechanism for all excitation energies}. Their characteristic parameters determined from Lorentzian line shape fitting yield $\omega_{\text{LO}}=\SI{681}{\per\cm}$ and an increased broadening and reduced asymmetry for higher LO scattering order $n$.
	\end{abstract}
	
	\maketitle
	
	\section{Introduction}\label{introduction}
	ScN is a transition metal nitride with rocksalt (rs) crystal structure (Fm$\bar{3}$m) as the only stable and wurtzite (wz) structure as a metastable phase \cite{Takeuchi.2002}. It is a promising material for certain applications due to its fundamental characteristics \cite{Biswas.2019}. For example, giant polarization charge densities were predicted for (111)rs-ScN/(0001)wz-GaN interfaces \cite{Adamski.2019} and an improvement of crystalline quality in epitaxial wz-GaN films with rs-ScN interlayers has been achieved \cite{Moram.2007}. Recently, the ternary alloy system wz-Sc\sub{x}Al\sub{1-x}N, which is stable up to 25-35\% AlN mole fraction \cite{Hoglund.2010,Satoh.2022}, attracts high research interest due to piezo- \cite{Ambacher.2021, Ambacher.2023} and ferroelectric properties \cite{Fichtner.2019,Wolff.2021} enabling e.g. wz-Sc\sub{x}Al\sub{1-x}N-barrier HEMT structures via PA-MBE growth \cite{Frei.2019}. As a counterpart to wz-AlN, rs-ScN is the binary base material for ternary rs-Al\sub{x}Sc\sub{1-x}N, which is under current investigation as well \cite{Satoh.2022,Deng.2015, Ambacher.2023} and stable up to $\approx$55\% ScN mole fraction \cite{Satoh.2022}. \\
	
	Already early theoretical studies \cite{Weinberger.1971,Neckel.1975} concluded that rs-ScN is an indirect semiconductor with its valence band maximum(VBM) located at the $\Gamma$-point and the conduction band minimum(CBM) located at the $X$-point of the Brillouin zone, which is confirmed also by more recent studies \cite{Gall.2001,Qteish.2006,Deng.2015b,Mu.2021}. Some of the first experimental attempts including crystal growth and optical characterization were performed by Dismukes \textit{et al.} \cite{Dismukes.1972}, Travaligni \textit{et al}. \cite{Travaglini.1986}, and by Gall \textit{et al}. \cite{Gall.1998,Gall.2001}. In the past, a very high unintentional free electron concentration usually masked the intrinsic properties of rs-ScN regardless of the growth technique. However, in the last 15 years there were new attempts to grow rs-ScN films with high crystal quality, mainly by halide/hydride vapor phase epitaxy (HVPE) \cite{Dismukes.1972,Oshima.2014}, sputter epitaxy \cite{Gregoire.2008,Saha.2013,Burmistrova.2013,Saha.2017} even at room temperature conditions \cite{Chowdhury.2022}, and molecular beam epitaxy (MBE) \cite{Smith.2001,Moram.2006b,Rao.2020}, going hand in hand with improved control of the unintentionally introduced free electrons. Sputter grown samples typically exhibit carrier densities higher than \SI{1E20}{\per\cm\cubed}  \cite{Gregoire.2008,Saha.2013,Deng.2015b}, MBE grown sample are reported to achieve \SI{5E18}{\per\cm\cubed}  \cite{AlBrithen.2004}. Oshima \textit{et al}. \cite{Oshima.2014} achieved record \SI{1.2E18}{\per\cm\cubed} by HVPE. The origin of the unintentional doping is attributed \tc{\chr}{to} nitrogen vacancies \cite{AlBrithen.2004,Kerdsongpanya.2012,Kumagai.2018} and oxygen impurities \cite{Moram.2008, Rowberg.2024}. It was already shown by Dismukes \textit{et al}. \cite{Dismukes.1972} and Oshima \textit{et al}. \cite{Oshima.2014} that the thickness of the rs-ScN layer has a massive impact on the structural quality and therefore dominantly determines the carrier concentration. These high unintentional carrier concentrations still remain one of the greatest challenges in growth and characterization of rs-ScN. \\
	
	\tc{\chr}{	
		For application in electrical and optical devices, knowledge of the optical constants of a material is crucial. Also, phonon and luminescence properties may yield relatively quick and non-destructive characterization of semiconductors. This work provides solid optical parameters, i.e. band gap values, phonon eigenfrequencies and broadenings, optical constants including precisely determined static and high frequency limits, and Born effective charges of quasi-bulk ScN. Knowledge of these parameter for a bulk-like single crystal sets a basis for understanding and design of thin films and heterostructures. Our results reveal several differences to state of the art ScN thin films, most striking the enhanced above-bandgap absorption, giant amplitude and narrow linewidth of the IR active phonon mode, and strong multiple phonon Raman scattering up to 6$th$ order. Those parameters partially stand in contrast to the properties of traditional III-V semiconductors.}
	
\section{experimental}\label{experimental}

For simplification we now use the notation ScN instead of rs-ScN. In this study we investigate a (100) oriented, $\approx$\SI{40}{\micro\m} thick ScN film deposited by HVPE on r-plane sapphire \cite{Oshima.2014}. X-ray diffraction (XRD) full widths at half maximum (FWHM) values are \SI{0.07}{\degree} and \SI{0.34}{\degree} for the (200) and the (131) reflexes, respectively, indicating relaxation and low dislocation densities. The free carrier concentration of our sample is determined by Hall effect measurements \tc{\chr}{with van-der-Pauw geometry} to be $n=\SI{1.2E18}{\per\cm\cubed}$ at room temperature, the corresponding mobility is \SI{264}{\cm\squared\per\V\per\s}. The origin of the free electrons is not clear, however it is much below the theoretical calculated degeneracy threshold of \tc{\chr}{$\approx\SI{3E19}{\per\cm\cubed}$} \cite{Mu.2021}. Therefore the sample represents one of the best ScN films currently available. We use optical characterization techniques, namely spectroscopic ellipsometry (SE) \tc{\chr}{in the infrared (IR) and near infrared (NIR) $-$ ultraviolett (UV) spectral range}, Raman spectroscopy [including photoluminescence (PL)], and \tc{\chr}{IR} reflectivity. Raman spectra were excited by laser wavelengths of either \SI{632.8}{\nm}, \SI{532.1}{\nm}, or \SI{472.9}{\nm}. A discussion of spectral resolutions and detailed experimental setups is reported in the supplement\cite{Supplement}. \\
	
	\begin{figure}[t]
		\centering
		\includegraphics[width=\linewidth]{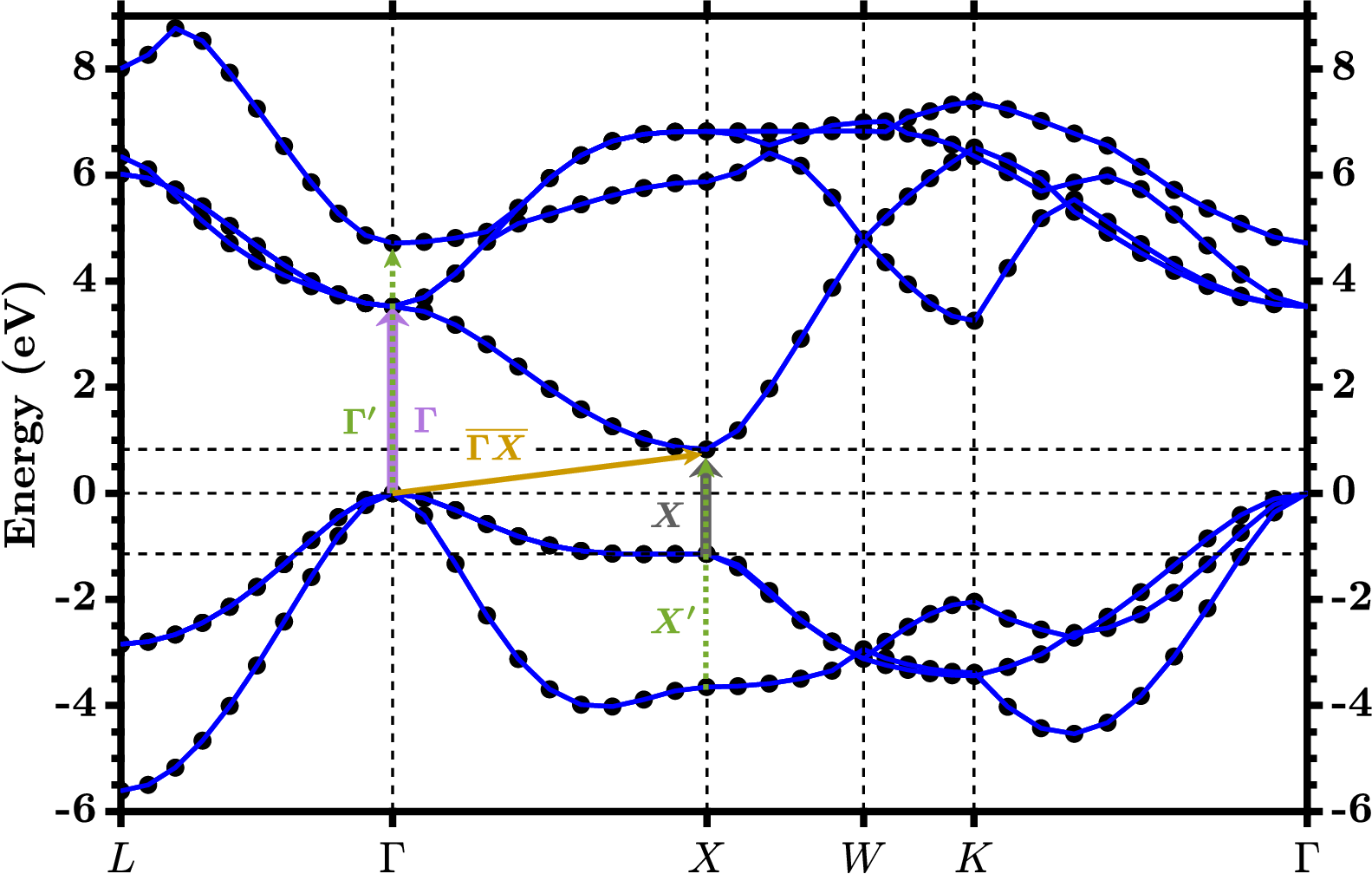}
		\caption{Numerically determined, exact-exchange-based quasiparticle bandstructure of rs-ScN from Ref.~\onlinecite{Qteish.2006} and main critical point transitions.}
		\label{BS Qteish}
	\end{figure}
	
	\section{theoretical models}
	The space group of rocksalt is $\mathrm{Fm\bar{3}m}$, the point group $\mathrm{\frac{4}{m}\bar{3}\frac{2}{m}}$. \tc{\chr}{Due 
		to the rocksalt structure with two atoms per unit cell
	}, only 6 phonon branches exist in ScN, where three of them are optical. The eigenfrequencies at the $\Gamma$-point are predicted to be \SI{632}{\per\cm}(\SI{78.4}{\meV}) for the longitudinal optical (LO) phonon and \SI{365}{\per\cm}(\SI{45.3}{\meV}) for the doubly degenerate transverse optical (TO) phonon mode \cite{Paudel.2009}. Experimental results suggest $2\omega_{\text{TO}}\approx \omega_{\text{LO}}$  [$\omega_{\text{LO}}=$\SI{686}{\per\cm} (\SI{85.1}{\meV}), $\omega_{\text{TO}}=$\SI{346}{\per\cm} (\SI{42.9}{\meV})] \cite{Uchiyama.2018} at the $\Gamma$-point, which constitutes an even larger LO-TO splitting than theoretically predicted \tc{\chr}{\cite{Paudel.2009}}. While first order Raman scattering in rs-structured crystals is symmetry forbidden, second and higher order processes are in general allowed \cite{Burstein.1965}, but exhibit a relatively weak scattering efficiency. One way to overcome the small efficiency of $n$th order phonon scattering is to choose an incident photon energy $E_\mathrm{i}$, for which \tc{\cho}{an outgoing} resonance condition is fulfilled, i.e. choosing an excitation energy \tc{\cho}{above the optical band gap ($E_{\text{Laser}}>E_{\mathrm{g},\text{opt}}$)}. The resonance enhancement of multiple phonon scattering in polar materials is expected to be particularly large for LO phonons due to the so-called Fröhlich mechanism \cite{Abrashev.1997}.\\

	Due to the non-vanishing \tc{\chr}{dipole-moment} under vibration of Sc\super{+}- and N\super{-}-ions the TO phonon mode is IR-active, exhibiting the symmetry $T_{\mathrm{1u}}$. Additionally, in the IR spectral region one has to take into account the free carrier absorption for the given carrier concentration ($n>\SI{E18}{\per\cm\cubed}$). For a single IR-active phonon mode and a Drude contribution the dielectric function for $\hbar\omega \ll E_{\mathrm{g},\text{opt}}$ is described by
	\begin{align}
		\varepsilon(\omega) = \varepsilon_{\infty} + \frac{S'\omega_{\text{TO}}^2}{\omega_{\text{TO}}^2-\omega^2-i\gamma_{\text{TO}}\omega} - \frac{\omega_{\mathrm{P}}^2}{\omega^2+i\gamma_{\mathrm{P}}\omega} \label{drude lorentz func}
	\end{align}
	where $S'=\varepsilon_\mathrm{stat}-\varepsilon_{\infty}$ and $\varepsilon_\mathrm{stat}$ /$\varepsilon_{\infty}$ are the low-/high-frequency limits of the $\varepsilon(\omega)$, respectively. Here, with the presence of a plasmon absorption, $\varepsilon_\mathrm{stat}$ is the limit $\varepsilon(\omega\to 0)$ without the Drude contribution. The plasma frequency $\omega_{\mathrm{P}}$ is given by
	\begin{align}
		\omega_{\mathrm{P}}^2 = \frac{ne^2}{\varepsilon_0m^*_{\mathrm{e},\text{opt}}} \label{plasma freq}
	\end{align}
	where $m^*_{\mathrm{e},\text{opt}}$ is the optical effective mass \cite{Baron.2019} and $n$ the concentration of free electrons in the conduction band. For ScN the approximation of a constant $\varepsilon_{\infty}$ does not hold satisfactorily, therefore we use the model by Shokhovets $et$ $al.$ \cite{Shokhovets.2003}:
	\begin{align}
		\varepsilon_{\text{vis}}(\omega) = 1+\frac{2}{\pi}\left[\frac{A_{\mathrm{g}}}{2}\ln{\left(\frac{E_{\mathrm{h}}^2-(\hbar\omega)^2}{E_{\mathrm{g}}^2-(\hbar\omega)^2}\right)}+\frac{A_{\mathrm{h}}E_\mathrm{h}}{E_{\mathrm{h}}^2-(\hbar\omega)^2}\right] \label{eps shokho}
	\end{align}
	We follow the notation of ref. \tc{\chr}{[\onlinecite{Feneberg.2016}]}. Within this model $\varepsilon_{\infty}$ is given by $\varepsilon_{\infty}=\varepsilon_{\text{vis}}(\omega\to 0)$ and $\varepsilon_{\infty}$ in eqn. (\ref{drude lorentz func}) is replaced by $\varepsilon_{\text{vis}}$ from eqn. (\ref{eps shokho}). Additionally, the LO-TO-splitting and $\varepsilon_{\infty}$ yield the Born effective charges $Z^*$, given by \cite{Zhang.2010, Gonze.1997}
	\begin{align}
		\omega_{\text{LO}}^2-\omega_{\text{TO}}^2 = \frac{(Z^*e)^2}{\varepsilon_0\varepsilon_{\infty} V_0\mu}
		\label{born charge}
	\end{align}
	where $Z^*=Z^*_{\mathrm{Sc}}=-Z^*_{\mathrm{N}}$ fulfills the required sum rule, $V_0$ is the primitive unit cell volume, and $\mu = M_{\text{Sc}}M_\text{N}/(M_{\text{Sc}}+M_\text{N})$ the reduced mass. The lattice constant of strain-free ScN is well known to be $a=\SI{4.505}{\angstrom}$ \cite{Moram.2006b}. \\
	
	%	\noindent\ff{Electronic transitions}\\
	\begin{figure*}[htb]
		\begin{subfigure}[b]{\sfs}
			\subcaption{}
			\includegraphics[width=\linewidth]{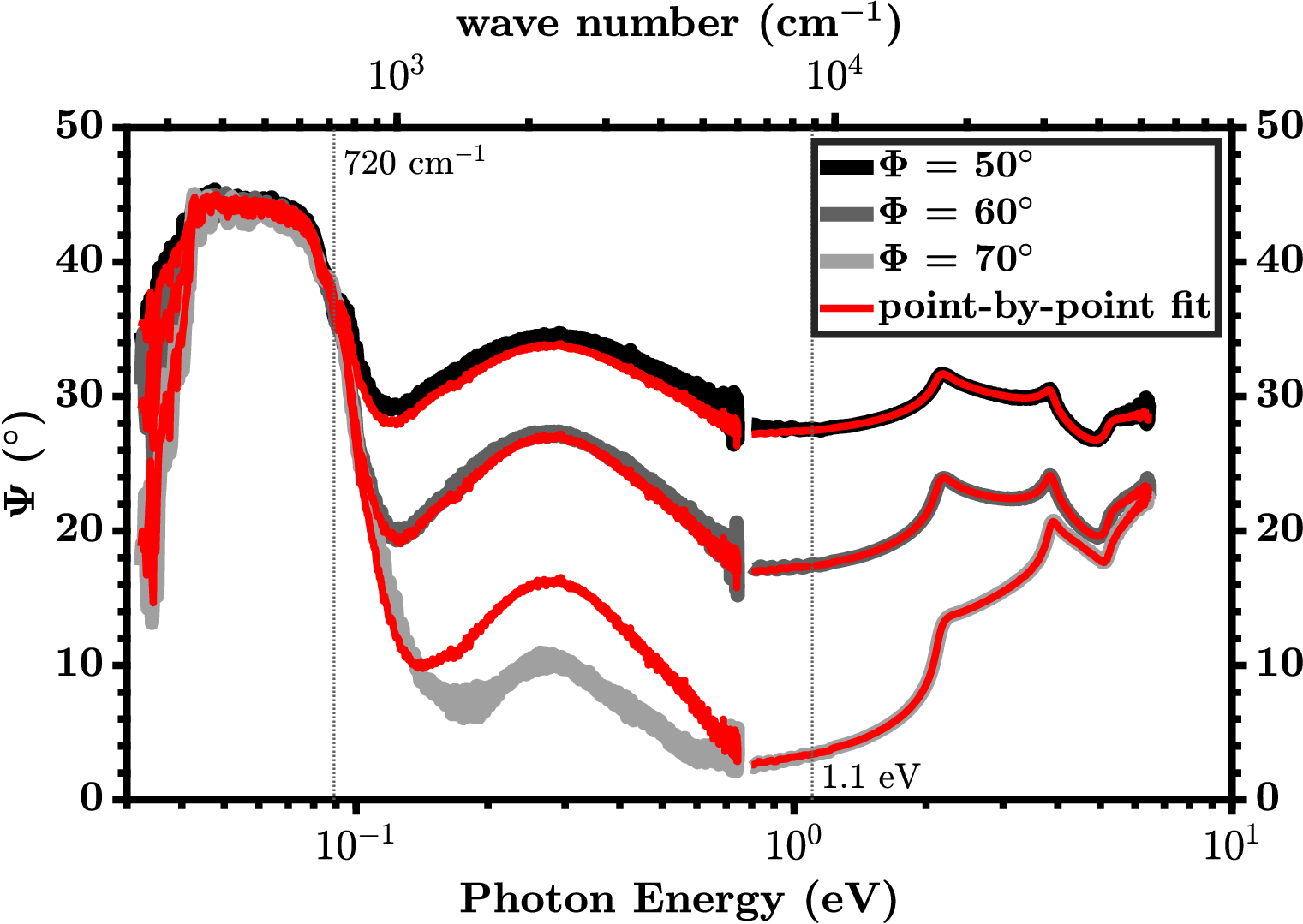}
			\label{Psi_pbp}
		\end{subfigure}	
		\hfil
		\begin{subfigure}[b]{\sfs}
			\subcaption{}
			\includegraphics[width=\linewidth]{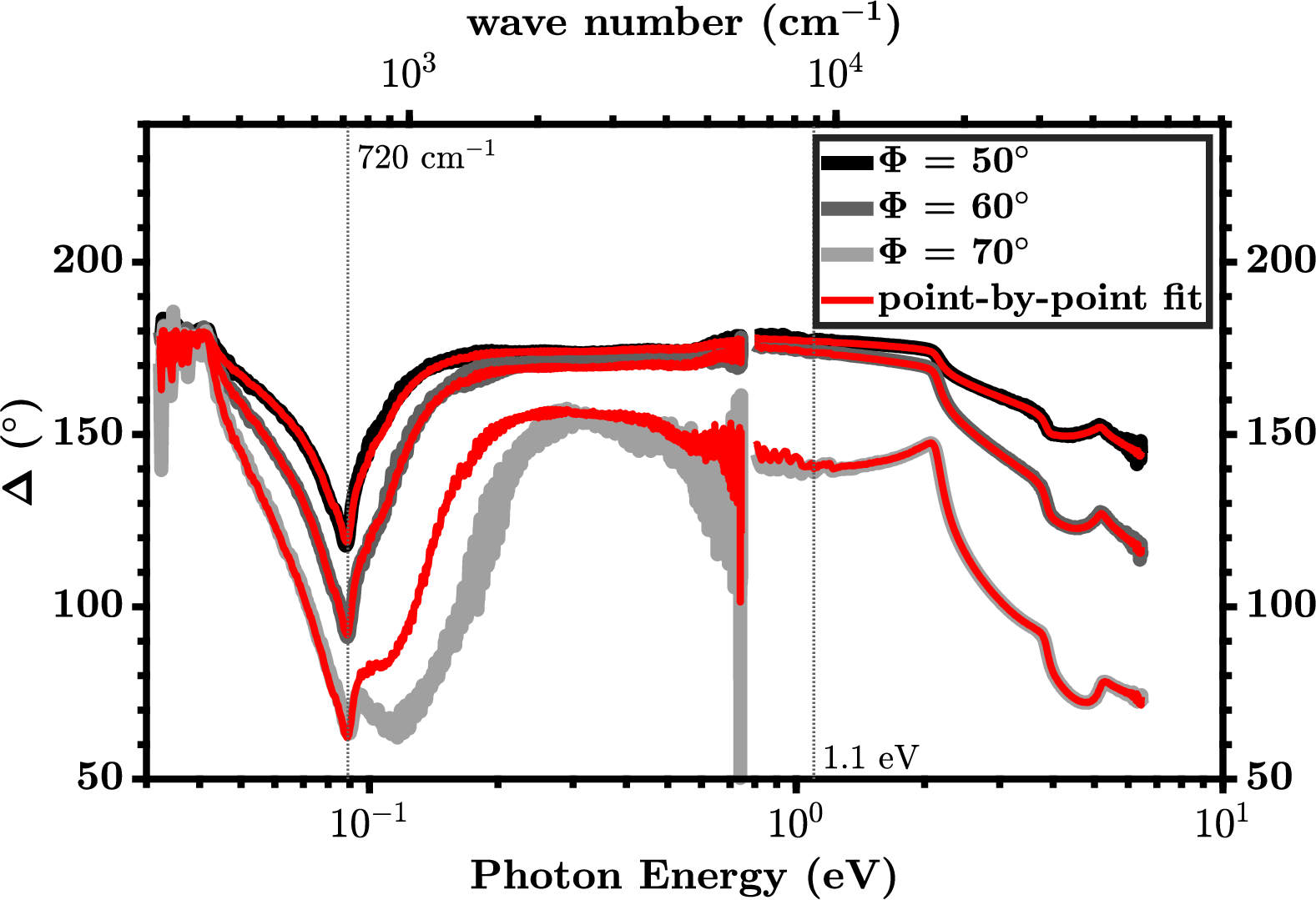}
			\label{Delta_pbp}
		\end{subfigure}	
		\hfil
		\begin{subfigure}[b]{\sfs}
			\subcaption{}
			\includegraphics[width=1\linewidth]{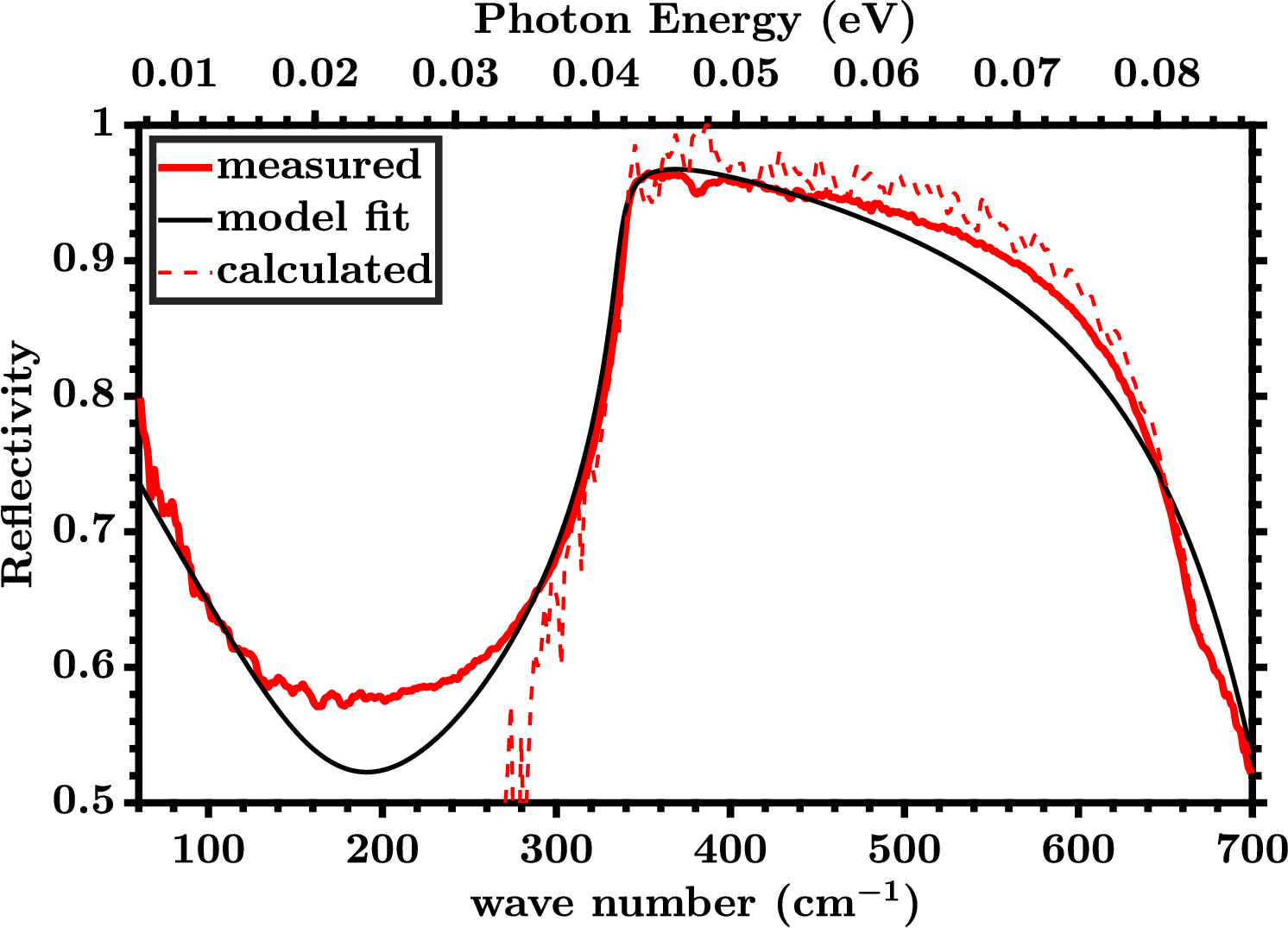}
			\label{reflectivity}
		\end{subfigure}
	\hfil
	\begin{subfigure}[b]{\sfs}
		\subcaption{}
		\includegraphics[width=\linewidth]{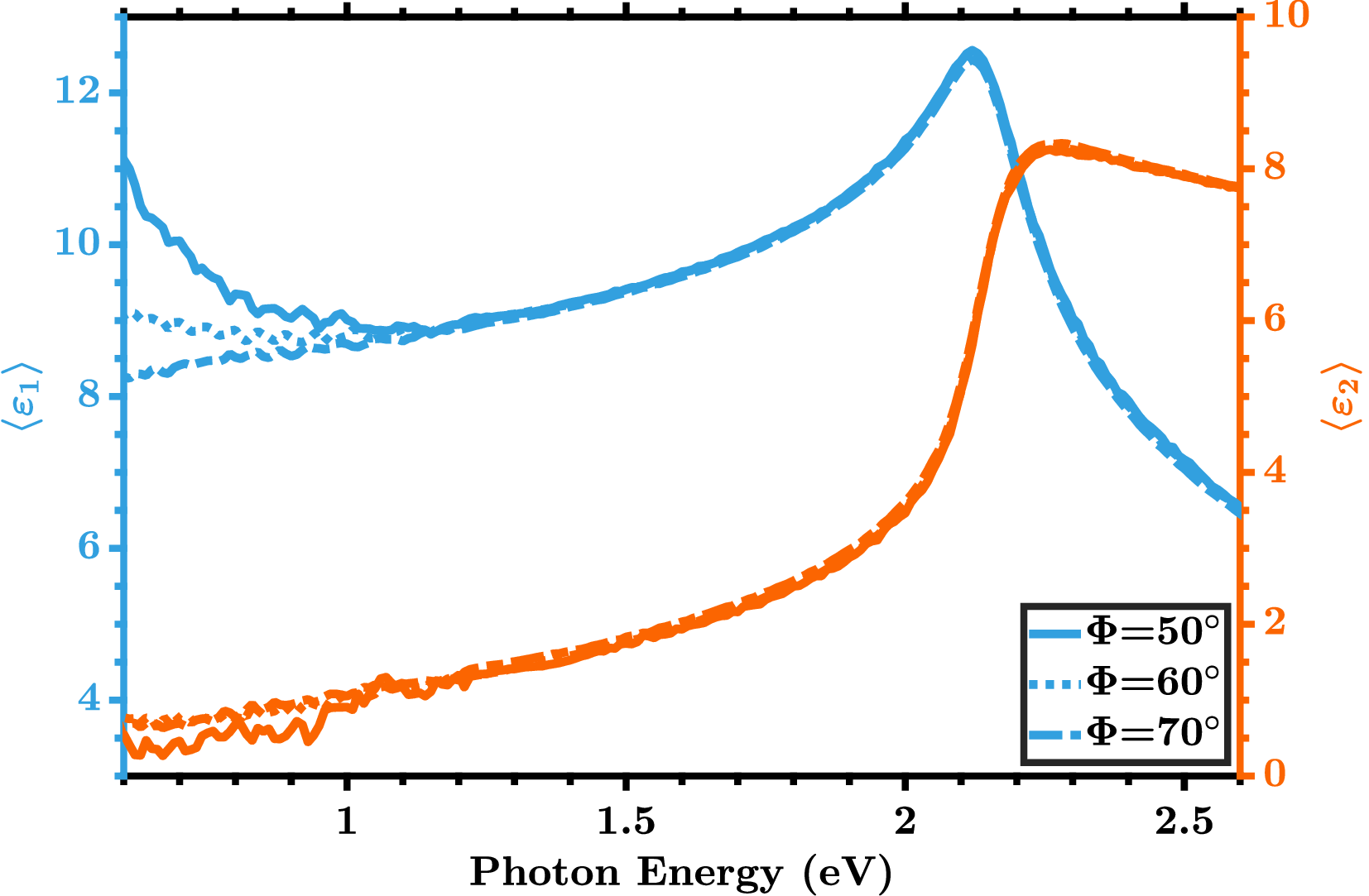}		
		\label{MN150R_pseudo_eps1_eps1}
	\end{subfigure}	
		\caption{Measured ellipsometric parameters (a) $\Psi$ and (b) $\Delta$ for three different incident angles (grey solid) and corresponding point-by-point dielectric function model fit results (red solid) for \tc{\chr}{IR-SE} and \tc{\chr}{UV-SE}. (c) measured IR reflectivity (red solid), model fit (black solid), and calculated reflectivity from point-by-point fitted $\varepsilon_1$ and $\varepsilon_2$  (red dashed) and (d) pseudo dielectric functions $\braket{\varepsilon_1}$ (blue) and $\braket{\varepsilon_2}$ (orange) from \tc{\chr}{UV-SE}.}
		\label{ellips_exp}
	\end{figure*}
	
	To describe the imaginary part of the dielectric function around the fundamental absorption edge we use the model by Elliott $et$ $al.$ \cite{Elliott.1957}. For the exciton continuum we have
	\begin{align}
		\varepsilon_2^{(\text{con})}(\omega) =	 \frac{C}{\hbar\omega}\frac{1+\text{erf}{\left[\frac{\hbar\omega-E_{\mathrm{g}}}{\gamma_{\mathrm{g}}}\right]}}{1-\exp{\left[-2\pi\sqrt{\vert\frac{E_{\mathrm{bX}}}{\hbar\omega-E_{\mathrm{g}}}\vert}\right]}}+\varepsilon_{\text{off}}
		\label{Elliott con}
	\end{align}
	where $E_{\mathrm{bX}}$ denotes the exciton binding energy, $\gamma_{\mathrm{g}}$ is the empirical width of the exciton continuum absorption, $C$ a constant taking into account the transition matrix element and $E_{\mathrm{g}}$ the bandgap energy \cite{Feneberg.2016}. The constant offset $\varepsilon_{\text{off}}$ is added to improve the fit results as discussed later. In addition, different electronic transitions are assigned to \tc{\chr}{van-Hove singularities} in the joint density of states (JDOS). From the electronic band structure shown in Fig. \ref{BS Qteish}, we can assume a $M_0$ critical point (CP) for the direct $X$-transition (at $\approx\SI{2}{\eV}$), a $M_1$ CP for the direct $\Gamma$-transition (at $\approx\SI{4}{\eV}$) and a $M_0$ CP for the direct $\Gamma'$- or $X'$-transition (at $\approx\SI{5.5}{\eV}$) to the second conduction- or valence band, respectively. For different types of CPs different line shapes of $\varepsilon(\omega)$ are expected. \\
	
	\section{results}\label{results Sec}
	
	\subsection{Ellipsometric angles and pseudo-dielectric function}
	
	The ellipsometric angles $\Psi$ and $\Delta$ shown in Figs.~\ref{Psi_pbp} and \ref{Delta_pbp} are converted to the so-called pseudo-dielectric function $\langle \varepsilon \rangle$, partially shown in Fig.~\ref{MN150R_pseudo_eps1_eps1}. \tc{\chr}{$\Psi$ and $\Delta$ are the polarization angle and phase shifts that emerge from the reflection of light with a distinct polarization at an interface. For a cubic crystal, they are linked to the diagonal Fresnel coefficients $r_{jj}$ by
		\begin{align}
		\rho = 	\frac{r_\mathrm{pp}}{r_\mathrm{ss}} = \tan(\Psi)e^{i\Delta}
		\end{align}
	and with to pseudo-dielectric function by
	\begin{align}
		\langle\varepsilon\rangle = \sin^2\Phi\left[1 + \tan^2\Phi\left(\frac{1-\rho}{1+\rho}\right)^2\right].
	\end{align}
	where $\Phi$ is the incident angle.} $\langle \varepsilon \rangle$ would be identical to $\varepsilon$ in case of an isotropic, semi-infinite, and perfectly smooth sample. \\

	For photon energies $<\SI{1.1\pm 0.1}{\eV}$ the data sets obtained with different angles of incidence do not merge to a single $\langle \varepsilon \rangle$. This threshold can be interpreted as the energy position, where incoherent interface reflections become relevant, i.e. the energy of the indirect band gap which is found between $\Gamma$ and $X$ points of the Brillouin zone. This indirect band gap energy of $\SI{1.1\pm 0.1}{\eV}$ is in good agreement with earlier results \cite{Gall.2001,Qteish.2006,Mu.2021,AlAtabi.2022}. Due to incoherent backside reflections in the transparent region [\SI{0.089}{\eV}(\SI{720}{\per\cm})$-$\SI{1.1}{\eV}]
	the point-by-point fit does not yield reliable results, which is obvious from Figs. \ref{Psi_pbp} and \ref{Delta_pbp}. Therefore we exclude this spectral range from the analysis of point-by-point fitted dielectric functions. The offset in $\Psi$ and $\Delta$ between \tc{\chr}{IR-SE} and \tc{\chr}{UV-SE} data is caused by different spot sizes.\\
	
	\begin{figure*}[t!]
		\centering
		\begin{subfigure}[b]{\sfs}
			\subcaption{}
			\includegraphics[width=\linewidth]{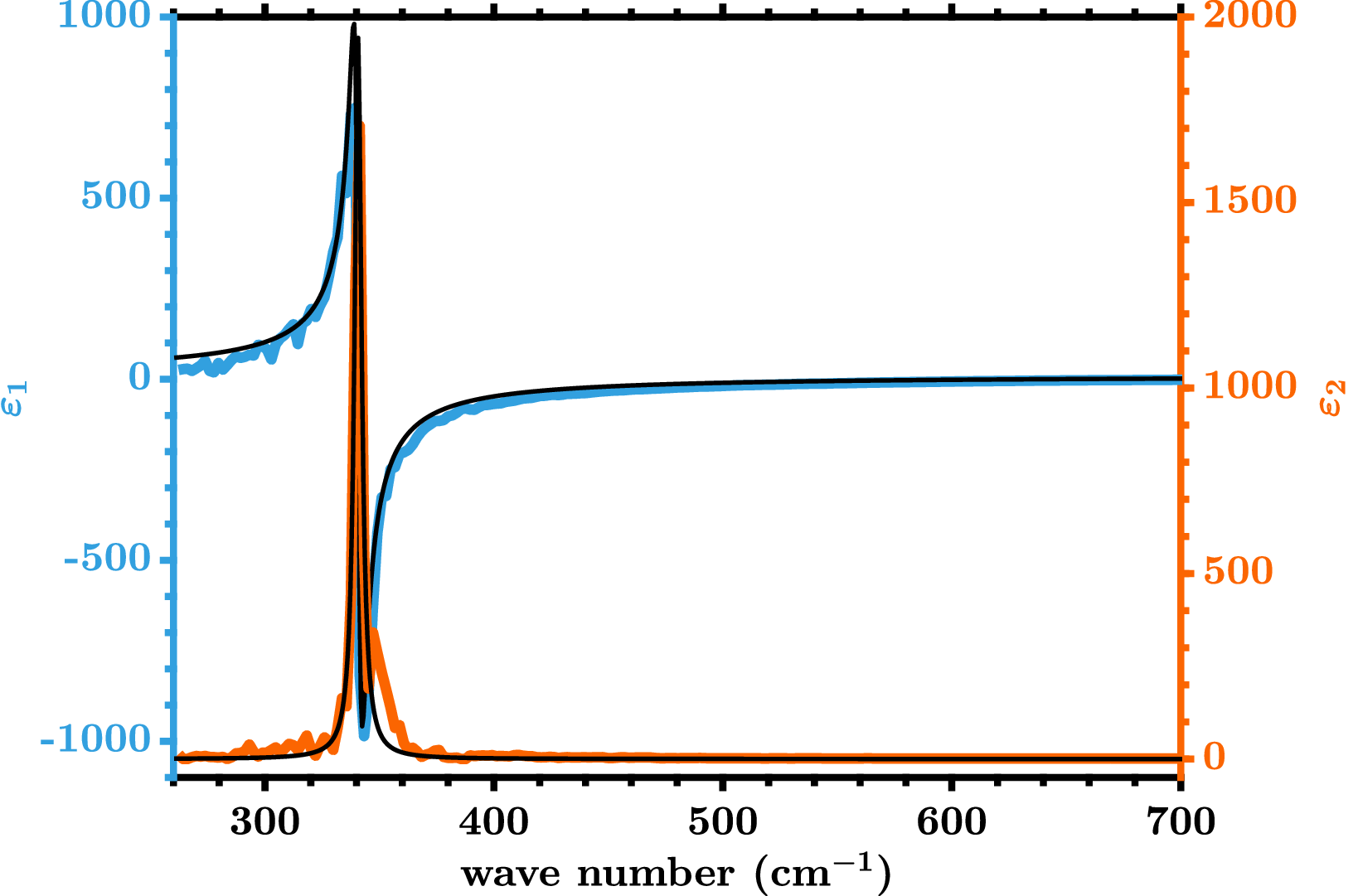}
			\label{IR DF refl}
		\end{subfigure}
		\hfil
		\begin{subfigure}[b]{\sfs}
			\subcaption{}
			\includegraphics[width=\linewidth]{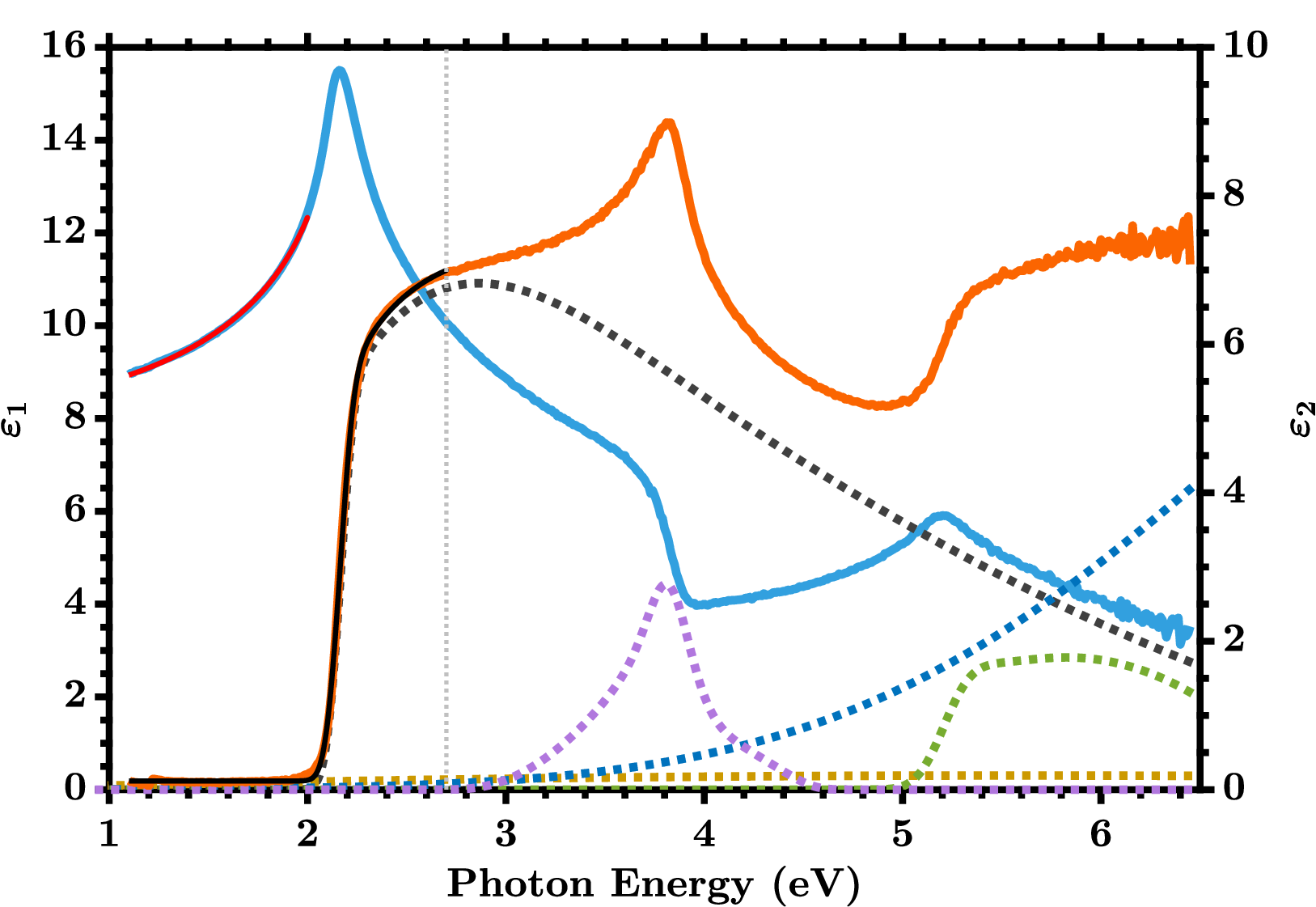}
			\label{eps1 eps2 + Elliott fit}
		\end{subfigure}
		\caption{(a) Real (blue) and imaginary (orange) part of the point-by-point fitted dielectric function derived from \tc{\chr}{IR-SE} and Lorentz-model fit (black lines). (b) Real (blue) and imaginary (orange) parts of the point-by-point fitted dielectric function derived from \tc{\chr}{UV-SE}, Elliott model fit to $\varepsilon_2$ (black line), Shokovets model fit to $\varepsilon_1$ (red line), and different model oscillators (colour dotted). The reference line indicates the upper bound (\SI{2.7}{\eV}) of the area, where only the direct $X$-transition exhibits $\varepsilon_2\gg0$.}
		\label{UV & IR DF}
	\end{figure*}	
	
\tc{\chr}{\subsection{NIR$-$UV dielectric function}}
	To obtain the dielectric function of ScN, we apply a simple bulk model without taking into account the substrate including the ScN layer and an obligatory roughness effective medium layer for the \tc{\chr}{NIR$-$UV} spectral range (\tc{\chr}{UV-SE}), which is \SI{8}{\nm} thick and has a layer/void ratio of 50\%. Point-by-point-fitting of the complex refractive index of the ScN layer yields the real and imaginary parts of the dielectric function, which are shown in Fig.~\ref{eps1 eps2 + Elliott fit} for the NIR-UV spectral range (point-by-point fit results are shown in Figs. \ref{Psi_pbp} and \ref{Delta_pbp}). This dielectric function is decomposed into different empirical functions (generalized parametric semiconductor oscillators provided by the analysis software WVASE32) exhibiting the typical characteristics of different critical point transitions. In Fig.~\ref{eps1 eps2 + Elliott fit} these contributions are shown as well. \\
	
	The characteristic energies of these \tc{\cho}{oscillators} are compared to the band structure calculated by Qteish \textit{et al.} \cite{Qteish.2006} (Fig.~\ref{BS Qteish}) and a very good qualitative agreement is found. The fundamental absorption edge of ScN is located at the $X$-point of the Brillouin-zone in agreement to Gall \textit{et al.} \cite{Gall.2001} and Haseman \textit{et al.} \cite{Haseman.2020} but in contrast to Saha \textit{et al.} \cite{Saha.2013}. The energetic ordering of the main transition features in the dielectric function (Fig.~\ref{eps1 eps2 + Elliott fit}) corroborates this fact: At $\approx\SI{2}{\eV}$ we find a $M_0$ like transition, at $\approx\SI{4}{\eV}$ a $M_1$ type transition, and again a $M_0$ type transition at $\approx\SI{5}{\eV}$. The band structure (Fig.~\ref{BS Qteish}) reveals the same ordering: the lowest direct CP transition occurs at the $X$-point ($dE_{\text{joint}}/dk>0$ for all directions, hence $M_0$ CP) and a second direct CP transition at the $\Gamma$-point ($dE_{\text{joint}}/dk<0$ only for $\overline{\Gamma X}$-direction, hence $M_1$ CP). The third CP transition could be assigned to two different points: (i) VB($\Gamma$) to second CB($\Gamma$) or (i) second VB($X$) to CB($X$). Possibly they both appear at around the same energy. The remaining contributions to our model are a general higher energy contribution (Fig. \ref{eps1 eps2 + Elliott fit}, blue dotted) and a small background like contribution describing the indirect $\overline{\Gamma X}$ transition (Fig. \ref{eps1 eps2 + Elliott fit}, yellow dotted). \\
	
	A quantitative analysis of the dielectric function for $E>\SI{1.1}{\eV}$ reveals the fundamental absorption edge at $E_{\mathrm{g}} =\SI{2.16}{\eV}$ with a width of $\gamma_{\mathrm{g}} = \SI{76}{\meV}$ from Elliott's model (eqn. (\ref{Elliott con})). Applying Shokhovets' model (eqn. \ref{eps shokho}) we obtain $A_{\mathrm{g}}=\SI{7.1}{}$ and $A_{\mathrm{h}}=\SI{27.4}{\eV}$ which precisely yields $\varepsilon_{\infty}=8.3$ ($E_{\mathrm{g}}=\SI{2.16}{\eV}$ and $E_{\mathrm{h}}=\SI{5.20}{eV}$ were fixed). The fitted curves and the real and imaginary parts of the dielectric function are displayed in Fig. \ref{eps1 eps2 + Elliott fit}. Shokhovets' model was applied only to $\varepsilon_1$ for $E<\SI{2}{\eV}$. Elliott's model is valid only for a single $M_0$ type VB to CB transition. Therefore, the fit was applied to the spectral region of $\varepsilon_2(\omega)$, where all other transitions have $\varepsilon_2\approx 0$, which is the case for $E<\SI{2.7}{\eV}$ (see Fig. \ref{eps1 eps2 + Elliott fit}). What still remains is the indirect absorption edge contribution, which we approximate by a constant offset $\varepsilon_{\text{off}}$ in eqn. (\ref{Elliott con}). Although we do not observe any discrete exciton state, we obtain the exciton binding energy $E_{\mathrm{bX}}=\SI{14}{\meV}$ as an estimate. For the transition energies $E_{\Gamma}$ and $E_{\Gamma'}$ we determine the inflection points of $\varepsilon_2$ to be $E_{\Gamma}=\SI{3.75}{\eV}$ and $E_{\Gamma'}=\SI{5.20}{\eV}$ in good agreement with the electronic band structure shown in Fig. \ref{BS Qteish}. \\
	
	Further, we observe a relatively high absolute $\varepsilon_2$ above the fundamental bandgap ($\varepsilon_2\approx 7$ at \SI{3}{\eV}) which yields an absorption coefficient $\alpha=\SI{3.4E5}{\per\cm}$ at \SI{3}{\eV}, which is significantly larger than that of e.g. GaN \cite{Muth.1997} or other common III-V nitrides at $\approx\SI{1}{\eV}$ above the absorption onset. Overall, our observations are similar to those of Dinh $et$ $al.$ \cite{Dinh.2023} qualitatively, but our absolute amplitudes of $\varepsilon_1$ and $\varepsilon_2$ are larger by a factor of $\approx 1.3-1.5$. Additionally, we can derive the Born effective charges from our $\varepsilon_{\infty}$ and later on determined $\omega_{\text{LO,TO}}$. We have \tc{\chm}{$Z^*_{\mathrm{Sc}}=-Z^*_{\mathrm{N}}=3.81$} (eqn. (\ref{born charge})), which is well above the expected ionization levels of 3/-3 for Sc/N respectively, but significantly lower than it was theoretically predicted \cite{Saha.2010}. This hints towards a strong bonding hybridization, leading to an admixture of N $2p$ and Sc $3d$ states in the highest occupied valence band, which was already shown by theoretical calculations \cite{Qteish.2006}. The same behaviour was observed for the related rs-CrN by both, theory \cite{Herwadkar.2009} and experiment($Z^*=4.4$)  \cite{Zhang.2010}.  \\
	
	\subsection{IR dielectric function}
	
	In the \tc{\chr}{IR} spectral range, a broad reststrahlenband arising from the phonon absorption and a free carrier absorption for $\omega<\SI{200}{\per\cm}$ is observed from IR reflectivity measurements (Fig. \ref{reflectivity}). From the model fit [eqns. (\ref{drude lorentz func}) and (\ref{eps shokho})] we determine Drude parameters of $\omega_{\mathrm{P}}=\SI{1200}{\per\cm}$ and $\gamma_{\mathrm{P}}=\SI{250}{\per\cm}$, yielding an effective electron mass of $m^*_\mathrm{e}=0.07m_\mathrm{e}$[eqn. (\ref{plasma freq})]. This effective electron mass is well below various theoretical calculated \cite{Mu.2021,Qteish.2006,Saha.2010} and experimentally determined \cite{Deng.2015b,AlAtabi.2022} values [$m_{\text{DOS}}\approx(0.35\pm 0.05)m_{\mathrm{e}}$]. In Fig. \ref{reflectivity} it is obvious, that the applied model does not match the data well, especially for wave numbers below $\SI{250}{\per\cm}$. Therefore, we conclude that the Drude formalism currently does not hold satisfactorily for free electrons in ScN and the determined plasmon parameters are not trustworthy. Evaluation of the \tc{\chr}{IR-SE} data (see Figs. \ref{Psi_pbp} and \ref{Delta_pbp}) yields more accurate results, but due to the limited spectral range, the plasmon absorption is not detectable.\\
	
	The point-by-point fitted dielectric function shown in Fig. \ref{IR DF refl} shows a strong phonon absorption with a maximum of $\varepsilon_2\approx 1700$ and the reflectivity calculated from this dielectric function matches the measured reflectivity well(Fig. \ref{reflectivity}). A similar value of $\varepsilon_2$ was reported before \cite{Maurya.2022} and hints towards a partially ionic bond in ScN, which was already theoretically predicted in 1971 by Weinberger $et$ $al.$ \cite{Weinberger.1971}. A Lorentz fit [eqn. (\ref{drude lorentz func}) without the Drude contribution] including our fixed $\varepsilon_{\text{vis}}(\omega)$ [eqn. (\ref{eps shokho})] from the \tc{\chr}{UV-SE} yields a TO-frequency of $\omega_{\text{TO}}=\SI{340.7}{\per\cm}$, broadening $\gamma_{\text{TO}}=\SI{3.7}{\per\cm}$, an amplitude $S=21.2$ and hence $\varepsilon_{\text{stat}}=29.5$. This broadening $\gamma_{\text{TO}}$ is about twice the selected spectral resolution (see supplement \cite{Supplement}). In comparison to common III-V nitride semiconductors such as GaN the oscillator strength of the TO phonon and hence $\varepsilon_{\text{stat}}$ is very high, and thus semiconducting ScN is a so called high-$k$ dielectric. The determined TO frequency ($\omega_{\text{TO}}=\SI{42.24}{\meV}$) matches well with the experimentally determined phonon dispersion of Uchiyama $et$ $al.$ ($\omega_{\text{TO}}(\Gamma+\delta k)=\SI{42.9}{\meV}$) \cite{Uchiyama.2018}, but is in contrast to several theoretical calculations \cite{Paudel.2009,Saha.2010} and other experiments \cite{Maurya.2022}. 
	
	\begin{figure*}[t]
		\begin{subfigure}[b]{\sfs}
			\subcaption{}
			\centering
			\includegraphics[width=\linewidth]{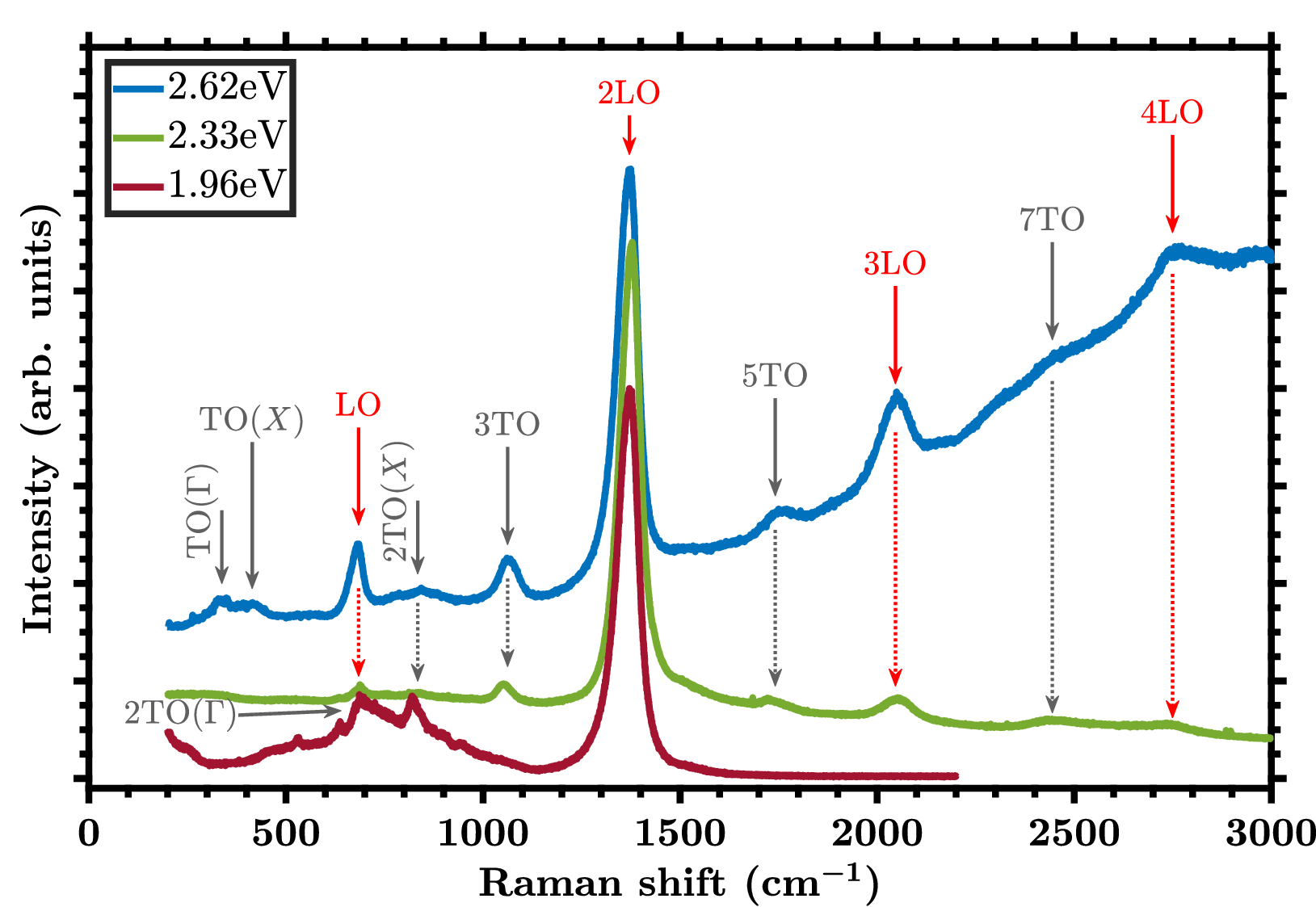}
			\label{Raman_comp_excitation}
		\end{subfigure}
		\hfil
		\begin{subfigure}[b]{\sfs}
			\subcaption{}
			\centering
			\includegraphics[width=\linewidth]{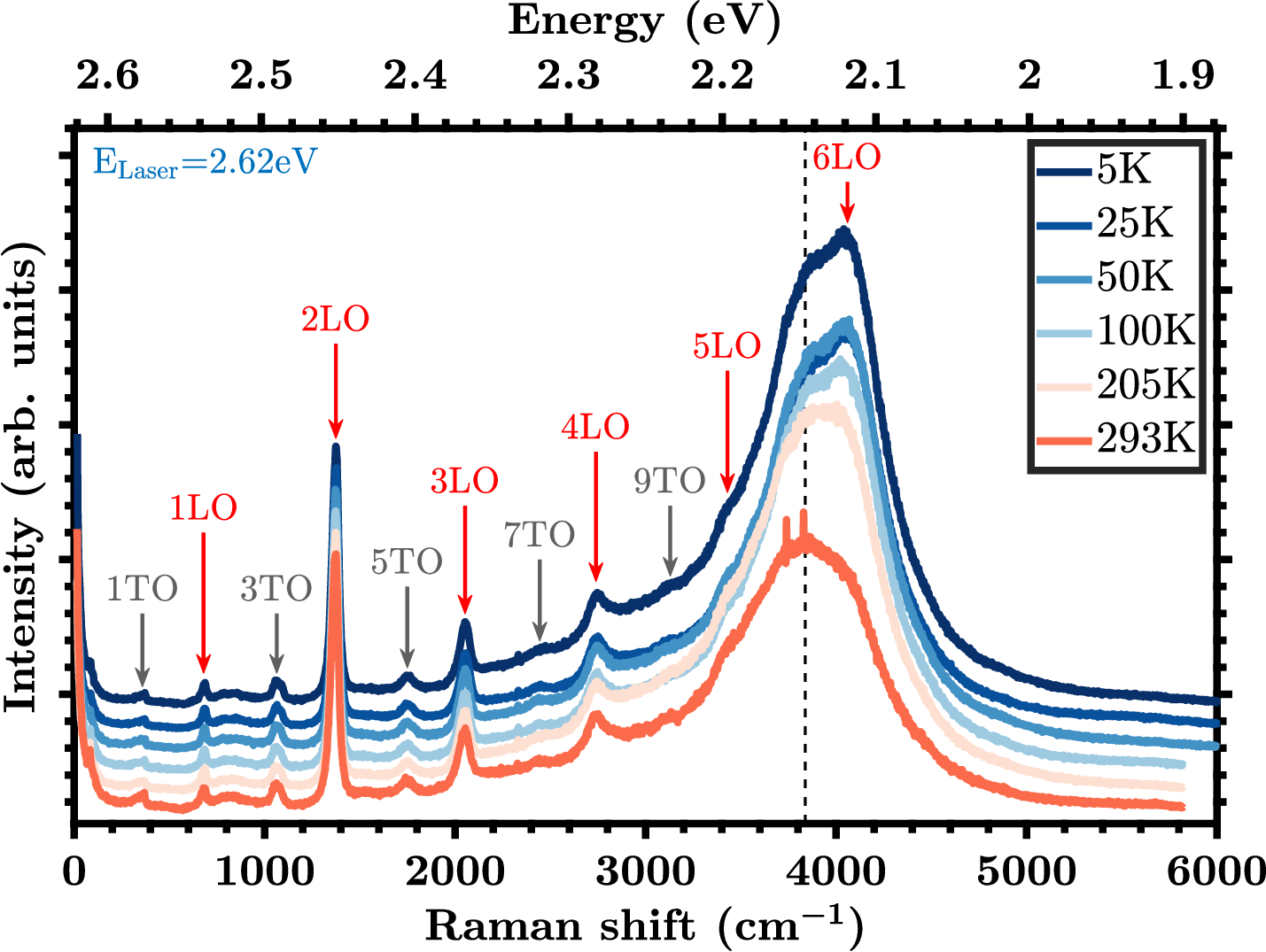}
			\label{Raman PL comp temp}
		\end{subfigure}
		\caption{(a) Room temperature Raman spectra at three different excitation wavelength and (b) Raman and PL spectra from \SI{5}{\K}$-$\SI{293}{\K} at an incident laser energy of \SI{2.62}{\eV}.}
	\end{figure*}	
	
	\subsection{Raman and PL spectra}
	
	For Raman and PL measurements we employed three different incident laser wavelengths: \SI{472.9}{\nm} (\SI{2.62}{\eV}), \SI{532.1}{\nm} (\SI{2.33}{\eV}), and \SI{632.8}{\nm} (\SI{1.96}{\eV}). In Fig. \ref{Raman_comp_excitation} the Raman spectra obtained with different excitation wavelengths are shown. These spectra are measured with a better resolution than photoluminescence (see supplement). For all three incident laser energies the allowed \cite{Burstein.1965} second-order scattering by two $\text{LO}$ phonons (2$\text{LO}$ peak) is the dominant scattering process with $\omega_{2\text{LO}}\approx\SI{1370}{\per\cm}$. Although first order scattering is Raman-forbidden, we observe weak first order Raman scattering for excitation energies of \SI{2.33}{\eV} and \SI{2.62}{\eV}. Assigning the multiple ($n$th order) phonon scattering peaks has to be done carefully for $\omega_{\text{LO}}\approx2\omega_{\text{TO}}$ and hence signals can be undistinguishable. The Raman spectrum at an incident laser energy of $\SI{2.62}{\eV}$ shows $n$LO scattering up to 4LO (see Fig. \ref{Raman_comp_excitation}, blue line). In the same figure, $n$TO signals are labeled for odd $n$. Due to $\omega_{\text{LO}}\approx2\omega_{\text{TO}}$ these Raman signals could either be $(2n-1)$TO or $n$LO+TO peaks. We observe two different first order TO related peaks at $\approx\SI{336}{\per\cm}$ and $\approx\SI{415}{\per\cm}$ with \SI{2.62}{\eV} excitation. They can be assigned to TO($\Gamma$) and TO($X$) in good agreement with previous experimentally determined values at $\Gamma$ ($\omega_{TO}=\SI{346}{\per\cm}$) and at $X$ ($\omega_{\tc{\chr}{\mathrm{TO}}}=\SI{408}{\per\cm}$) \tc{\cho}{points of the Brillouin zone} \cite{Uchiyama.2018}. \\
	
	Changing the laser energy to $\SI{2.33}{\eV}$ yields $E_{\text{Laser}}\approx E_{\mathrm{g},\text{opt}}+2E_{\text{LO}}$ and hence nearly perfect critical point resonance for the 2LO (and 4TO) scattering. Therefore, all other signals exhibit relatively small intensities. For excitation in the transparent region at \SI{1.96}{\eV}, the condition for the enhancement of multiple order phonon scattering by outgoing resonances is not fulfilled anymore. Indeed, while we still observe a dominant 2LO peak, no higher order scattering is found anymore (see Fig. \ref{Raman_comp_excitation}, red line). Additionally, we observe a broad contribution covering various narrower signals around $\SI{700}{\per\cm}$. The origin of this broad contribution is not known yet. Because we do not expect first order scattering \tc{\cho}{for the non-resonant excitation at \SI{1.96}{\eV}}, we tentatively assign the two features at $\approx\SI{690}{\per\cm}$ and $\approx\SI{830}{\per\cm}$ as 2TO($\Gamma$) and 2TO($X$) rather than first order LO scattering. The 2TO($X$) is also weakly visible at other incident laser energies. \\
	
\tc{\chm}{For a quantitative analysis, line shape fits are performed using standard Lorentzian type functions}. \tc{\chr}{Due to the luminescence signal, which is located directly at the position of the 2LO line, the spectrum for $E_\mathrm{Laser}=\SI{2.33}{\eV}$ is impossible to fit unambiguously and with \SI{1.96}{\eV} excitation we observe only second order scattering. Therefore, we choose to model only data recorded with $E_\mathrm{Laser}=\SI{2.62}{\eV}$.} Results are summarized in Tab. \ref{Raman_results} and fit details are reported in the supplement \cite{Supplement}. \tc{\chm}{From Tab. \ref{Raman_results} the dependence of the $n$-LO broadening $\gamma$ on the LO-order $n$ is obvious. We observe increasing broadening with higher scattering order $n$, while the broadening values for the $n$TO lines scatter around $\approx\SI{40}{}-\SI{100}{\per\cm}$.} An important result is the LO eigenfrequency obtained from fitting the first order LO signal. We have \tc{\chm}{$\omega_{\text{LO}}=\SI{684.5}{\per\cm}$} in good agreement with earlier experiments ($\omega_{\text{LO}}=\SI{686}{\per\cm}$) \cite{Uchiyama.2018,Maurya.2022}, but in contrast to theoretical calculations ($\omega_{\tc{\chr}{\mathrm{LO}}}\approx\SI{630}{\per\cm}$) \cite{Paudel.2009,Saha.2010} at the $\Gamma$-point. \tc{\chm}{The phonon lifetime can easily be derived as the inverse phonon broadening as $\tau_\mathrm{LO}=(2\pi\gamma_\mathrm{LO})^{-1}$. We have $\tau_\mathrm{LO}=\SI{0.2}{\ps}$, which is again in perfect agreement with previous experimental results ($\tau_\mathrm{LO}=\SI{0.21}{\ps}$ \cite{Uchiyama.2018}).} So in contrast to Dinh $et$ $al.$ \cite{Dinh.2023}, we rather conclude that the LO scattering takes place at the $\Gamma$-point due to both, momentum conservation and above discussed previous XRD results \cite{Uchiyama.2018}. The multiple \tc{\chm}{odd number} $\text{TO}$ scattering lines correspond to $\text{TO}$ frequencies of $\approx\SI{350}{\per\cm}$, which is between the \tc{\cho}{frequencies of the first-order TO phonon lines (see Tab. \ref{Raman_results}) indicating their origins as $i\mathrm{TO}(\Gamma)+j\mathrm{TO}(X)$ ($i+j = n$) combination modes}. \tc{\chm}{The multiple even number TO lines correspond to TO frequencies of $\approx\SI{337}{\per\cm}$, which indicates their origin as $n$TO($\Gamma$) lines.} \\
	
	\begin{table*}[t]
		\setlength{\tabcolsep}{1.2pt}
		\renewcommand{\arraystretch}{1.5}
		\centering
		\caption{\tc{\chm}{Characterization of single and multiple phonon resonant Raman scattering for an incident laser energy of \SI{2.62}{\eV}. Eigenfrequencies $\omega_0/n$, broadening parameters $\gamma_0$, and signal-to-noise ratios $\text{snr}$ were determined  by Lorentzian line shape fitting.}}
		\tc{\chm}{
		\begin{tabular}{cccccccccccccc} \hline
			& TO($\Gamma$) & TO($X$) & 2TO & LO & 3TO & 4TO & 2LO & 5TO & 6TO & 3LO & 7TO & 8TO & 4LO \\ \hline
			% $A$     & 0.044 & 0.030 & 0.142 & 0.092 & % 0.902 & 0.040 & 0.169 & 0.019 & 0.115 \\
			$\frac{\omega_0}{n}$ (\SI{}{\per\cm})  & 336$\pm$2 & 415$\pm$2 & 334.5$\pm$2 & 684.5$\pm$2 & 355.0$\pm$0.2 & 338.8$\pm$0.8 & 689.0$\pm$0.5 & 350.0$\pm$0.4 & 337.5$\pm$0.8 & 686$\pm$1 & 349.4$\pm$0.7 & 342.0$\pm$0.6 & 689$\pm$0.8 \\
			$\gamma$ (\SI{}{\per\cm})& 76$\pm$6  & 64$\pm$8  & 40$\pm$7 & 27$\pm$4  & 60$\pm$3  & 59$\pm$4 & 36$\pm$4 & 94$\pm$8 & 75$\pm$8 & 65$\pm$6 & 100$\pm$30 & 30$\pm$20 & 210$\pm$20 \\
			$\text{snr}$   & 6.3$\pm$0.3 & 4.0$\pm$0.3 & 9$\pm$3 & 14$\pm$4 & 12.8$\pm$0.3  & 66$\pm$8 & 63$\pm$9  & 6.5$\pm$0.3 & 8$\pm$2  & 10$\pm$2  & 1.3$\pm$0.2 & 0.6$\pm$0.3  & 5.3$\pm$0.4 \\ \hline
		\end{tabular} }
		\label{Raman_results}
	\end{table*}
	
	To investigate the photoluminescence signal, we scan a larger spectral range and apply lower temperatures down to \SI{5}{\K}. These spectra demonstrate multi-phonon scattering and luminescence in a single measurement, but do not allow precise evaluation of the $n$-phonon peak positions due to the decreased accuracy (see supplement \cite{Supplement}). In Fig. \ref{Raman PL comp temp} the Raman/PL spectra are presented from \SI{1.9}{}-\SI{2.6}{\eV} for different temperatures. We find a broad luminescence peak with a maximum at around $\approx$\SI{2.15}{\eV} and multi-phonon Raman signals up to 6LO and 9TO. Surprisingly, the center energy of the luminescence signal seems to decrease \tc{\chr}{slightly} with decreasing temperature. \tc{\chr}{Accordingly, the 6LO Raman line exhibits a much larger intensity with respect to the lower order modes at low temperature (\SI{5}{\K}). This enhancement is caused by a better matching of the resonance condition resulting from the slight decrease of the direct band gap.}  Comparing the luminescence peak to our ellipsometry results we conclude that luminescence arises from direct band-to-band recombination at the $X$-point. Concerning the band structure shown in Fig. \ref{BS Qteish}, the VB seems to be only a saddle point and photo-generated holes could easily thermalize towards $\Gamma$. Our observation thus strongly indicates a local VB maximum at the $X$-point. For a degenerately doped ScN single crystal ($n_{\text{Hall}}=$\SI{2.2E21}{\per\cm\cubed}) Al Atabi $et$ $al.$ \cite{AlAtabi.2022} already reported ARPES measurements indicating such a local VB maximum at the $X$-point. \\
	
	\begin{table}[t]
		\setlength{\tabcolsep}{8pt}
		\renewcommand{\arraystretch}{1.5}
		\centering
		\caption{Overview of results. Note that $\omega_{\text{LO}}$ displays the eigenfrequency of the first order LO signal at $E_{\text{Laser}}=\SI{2.62}{\eV}$. For details about the experimental accuracy see supplement \cite{Supplement}.}
		\begin{tabular}{cccc} \hline
			parameter & unit & value & accuracy \\ \hline
			$\omega_{\text{TO}}$  & \SI{}{\per\cm} & 340.7 & $\pm$0.03 \\ 
			$\gamma_{\text{TO}}$ & \SI{}{\per\cm}  & 3.7 & $\pm$0.06 \\
			$\varepsilon_{\text{stat}}$ & & 29.5 & $\pm$0.5 \\
			$\varepsilon_{\infty}$ & & 8.3 & $\pm$0.2\\
		    $\omega_{\text{LO}}$ & \SI{}{\per\cm} & \tc{\chm}{684.5} & \tc{\chm}{$\pm$2} \\
			$\gamma_{\text{LO}}$ \tc{\chm}{($\tau_\mathrm{LO}$)} &  \SI{}{\per\cm} \tc{\chm}{(\SI{}{\ps})}  & \tc{\chm}{27 (0.20)} & \tc{\chm}{$\pm$4 ($\pm$0.03)} \\
			$\frac{\omega_{\text{LO}}^2}{\omega_{\text{TO}}^2}\frac{\varepsilon_{\infty}}{\varepsilon_{\text{stat}}}$ & & \tc{\chm}{1.14} & \tc{\chm}{$\pm$0.06} \\
			 $\omega_{\text{LO}}^{(\text{LST})}$ & \SI{}{\per\cm}  & 640 & $\pm$20  \\
			$Z^*_{\mathrm{Sc}}=-Z^*_{\mathrm{N}}$ &  & \tc{\chm}{3.81} & \tc{\chm}{$\pm$0.05} \\
			$Z^{*(\text{LST})}_{\mathrm{Sc}}=-Z^{*(\text{LST})}_{\mathrm{N}}$ &  & 3.50 & $\pm$0.2 \\
			$E_{\mathrm{bX}}$ & \SI{}{\meV} & 14 & $\pm$1 \\	 
			$E_{\overline{\Gamma X}}$ & $\SI{}{\eV}$ & 1.1& $\pm$0.1 \\
			$E_{\mathrm{g}}$ & \SI{}{\eV} & 2.16 & $\pm$0.002 \\
			$\gamma_{\mathrm{g}}$ & \SI{}{\meV} & 76 & $\pm$3 \\
			$E_{\Gamma}$ & \SI{}{\eV} & 3.75 & $\pm$0.02\\
			$E_{\Gamma'}$ & \SI{}{\eV} & 5.20 & $\pm$0.03\\
			\hline
		\end{tabular}
		\label{results}
	\end{table}

\section{Conclusions}
	
In summary, we investigated a HVPE grown quasi-bulk rs-ScN single crystal and derived fundamental material parameters which are listed in Tab. \ref{results} (see supplement for error analysis \cite{Supplement}). \tc{\chr}{UV-SE} precisely yields an absorption edge of $E_g=\SI{2.16}{\eV}$ from an Elliott model fit as well as $\varepsilon_{\infty}=8.3$ from Shokhovets model fit. Additionally, an estimate of the exciton binding energy of $E_{\mathrm{bX}}=\SI{14}{\meV}$ was extracted. \tc{\chr}{IR-SE} and IR reflectivity showed a strong phonon and a weak free carrier (plasmon) absorption. Drude-Lorentz fitting yields $\omega_{\text{TO}}=\SI{340.7}{\per\cm}$, $\gamma_{\text{TO}}=\SI{3.7}{\per\cm}$, and $\varepsilon_{\text{stat}}=29.5$. \\
 
Checking the Lyddane-Sachs-Teller (LST) relation yields \tc{\chm}{$\frac{\omega_{LO}^2}{\omega_{TO}^2}\frac{\varepsilon_{\infty}}{\varepsilon_{\text{stat}}}=1.14\pm0.06$}, which is significantly off the expected value of 1. Deriving the LO frequency from LST yields $\omega_{LO}^{(\text{LST})}=\SI{640\pm20}{\per\cm}$, which is much closer to theoretically calculated values than our Raman result. Possible reasons for this deviation are the large LO broadening, which is neglected in the LST relation, or the coupling of the LO phonon mode with free charge carriers. Born effective charges are calculated as \tc{\chm}{$Z^*_\text{Sc} = 3.81$} and $Z^{*(\text{LST})}_\text{Sc} = 3.5$ using $\omega_{\text{LO}}$ and $\omega_{LO}^{(\text{LST})}$, respectively.\\

The free carrier absorption could not be described adequately by the Drude model. From Raman scattering experiments at different excitation wavelengths we observed efficient multi-phonon scattering up to 4LO and 7TO, and even 6LO and 9TO from temperature dependent combined Raman/PL spectroscopy. The $n$LO and $n$TO positions and broadening parameters are determined from Lorentzian line shape fits, which describe the spectra adequately for $E_{\text{Laser}}=\SI{2.62}{\eV}$. The LO broadening parameter increases with higher scattering order $n$ and the determined first order LO-frequency is \tc{\chm}{$\omega_{\text{LO}}=\SI{684.5}{\per\cm}$} in good agreement with previous experiments \cite{Uchiyama.2018,Maurya.2022}. In a combined Raman and PL measurement we observed a broad luminescence signal at $\approx\SI{2.15}{\eV}$ which perfectly matches the \tc{\chr}{UV-SE} determined direct bandgap of $\SI{2.16}{\eV}$. From this we assume a possible local valence band maximum at the $X$-point in agreement with earlier ARPES measurements \cite{AlAtabi.2022}.\\
	
\section{Acknowledgement}
This work was funded by the DFG within the framework of  the priority programme 2312 (GaNius) under project FE 1453/2-1. We acknowledge the support by the project Quantum materials for applications in sustainable technologies, CZ.02.01.01/00/22\textbackslash\textunderscore 008/0004572, the Czech Science Foundation (GACR) under Project No. GA20-10377S and the CzechNanoLab Research Infrastructure supported by MEYS CR (LM2023051). We thank Patrick Rinke for providing the DFT data.
	
	\bibliography{ScN_quasi_bulk_opt_props.bib}
	
\end{document}

\title{Supplement: Band gaps and phonons of quasi-bulk rocksalt ScN}
	\author{Jona Grümbel}\email{jona.gruembel@ovgu.de}  
	\affiliation{Institut für Physik, Otto-von-Guericke-Universität Magdeburg, Universitätsplatz 2, 39106, Magdeburg, Germany}
	\author{Yuichi Oshima}\affiliation{Research Center for Electronic and Optical Materials, National Institute for Materials Science , 1-1 Namiki, Tsukuba, Ibaraki 305-0044, Japan}
	\author{Adam Dubroka} \affiliation{Department of Condensed Matter Physics, Masaryk University, Kotl\'a\v{r}sk\'a 2, 611-37 Brno, Czech Republic}
	\author{Manfred Ramsteiner} \affiliation{ Paul-Drude-Institut für Festkörperelektronik (PDI), Hausvogteiplatz 5-7, 10117, Berlin, Germany}
	\author{Rüdiger Goldhahn}\affiliation{Institut für Physik, Otto-von-Guericke-Universität Magdeburg, Universitätsplatz 2, 39106, Magdeburg, Germany}
	\author{Martin Feneberg}\affiliation{Institut für Physik, Otto-von-Guericke-Universität Magdeburg, Universitätsplatz 2, 39106, Magdeburg, Germany}
	
	\maketitle

	\section*{SI. experimental setups}
	For our ellipsometry/reflectivity measurements we use three different machines, where the WVASE32 ellipsometry software (Woollam) was used for both data recording and analyzing:
	\begin{enumerate}
		\item[(I)]  A Wollam VASE rotating compensator UV ellipsometer for \SI{0.5}{}$-$\SI{6.5}{\eV} with a HS190 monochromator and a high pressure Xe-lamp as light source. The spectral resolution trace is shown in Fig. \ref{Resolution UV ellips}.
		\item[(II)] A Woollam VASE fourier transform IR ellipsometer for \SI{280}{}$-$\SI{6000}{\per\cm} (\SI{35}{}$-$\SI{750}{\meV}) with a Carbon Globar driven Michelson-Morley Interferometer as light source. The spectral resolution (energy step width) is set to \SI{2}{\per\cm}.
		\item[(III)] A Bruker Vertex 70V spectrometer for \SI{60}{}$-$\SI{700}{\per\cm}, equipped with a Hg lamp as source and DTGS detector. The spectral resolution is set to \SI{1}{\per\cm}.
	\end{enumerate} 
	Raman measurements are recorded with two different machines:
	\begin{enumerate}
		\item[(I)] A Tri-Vista Raman microscope with an incident laser wavelength of \SI{532.1}{\nm}, an Olympus BX50 microscope with a x50 objective for focussing, a Princeton Instruments SP2750i monochromator with a focal length of \SI{750}{\mm}, grating parameter of \SI{1800}{\per\mm}, slit width of \SI{200}{\micro\m} and a peltier-cooled charge coupled device (CCD) camera (2048px horizontal).
		\item[(II)]  A Horiba LabRam HR evolution system with incident laser wavelength of either \SI{632.8}{\nm}, or \SI{472.9}{\nm}, a x50 objective for focussing, a focal length of \SI{800}{\mm}, grating parameters of \SI{1800}{\per\mm} or \SI{600}{\per\mm}, slit width of \SI{200}{\micro\m} and a lN\sub{2}-cooled Symphony CCD camera (1024px horizontal).
	\end{enumerate}
	Different setups used to record Raman/PL spectra result in the following spectral resolutions $\Delta k_{\text{rel}}$ at \SI{300}{\per\cm} ($\Delta\lambda \approx \text{slit width}/(\text{grating}\times\text{focal length})$):
	\begin{enumerate}
		\item[Setup 1:] $\SI{633}{\nm}$, grating \SI{1800}{\per\mm}: $\Delta k_{\text{rel}}=\SI{3.3}{\per\cm}$
		\item[Setup 2:] $\SI{532}{\nm}$, grating \SI{1800}{\per\mm}: $\Delta k_{\text{rel}}=\SI{4.8}{\per\cm}$
		\item[Setup 3:] $\SI{473}{\nm}$, grating \SI{1800}{\per\mm}: $\Delta k_{\text{rel}}=\SI{6.0}{\per\cm}$
		\item[Setup 4:] $\SI{473}{\nm}$, grating \SI{600}{\per\mm}: $\Delta k_{\text{rel}}=\SI{18}{\per\cm}$ (\SI{2}{\meV})
	\end{enumerate}
	Setups 1-3 are used only for Raman measurements (Fig. 4a of the main article) and Setup 4 for the combined Raman/PL measurement (Fig. 4b of the main article).\\

\newpage

\begin{comment}
	\begin{figure*}[h]
		\begin{subfigure}[b]{\hsfs}
			\subcaption{}
			\includegraphics[width=\linewidth]{ScN/Ellipsometry/ScN_MN150R_IR_Psi_pbp}		
			\label{MN150R_IR Psi}
		\end{subfigure}	
		\hfil
		\begin{subfigure}[b]{\hsfs}
			\subcaption{}
			\includegraphics[width=\linewidth]{ScN/Ellipsometry/ScN_MN150R_IR_Delta_pbp}	
			\label{MN150R_IR_Delta}
		\end{subfigure}	
		\hfil
		\begin{subfigure}[b]{\hsfs}
			\subcaption{}
			\includegraphics[width=\linewidth]{ScN/Ellipsometry/ScN_MN150R_UV_Psi_pbp}		
			\label{MN150R_UV_Psi}
		\end{subfigure}
		\hfil
		\begin{subfigure}[b]{\hsfs}
			\subcaption{}
			\includegraphics[width=\linewidth]{ScN/Ellipsometry/ScN_MN150R_UV_Delta_pbp}		
			\label{MN150R_UV_Delta}
		\end{subfigure}	
		\caption{Measured ellipsometric parameters (a) IRSE $\Psi$, (b) IRSE $\Delta$, (c) UVSE $\Psi$, (d) UVSE $\Delta$, and corresponding point-by-point fit results excluding additional reflectivity measurements in the IRSE spectral range.}
		\label{ellips data}
	\end{figure*}
\end{comment}

\section*{SII. ellipsometry procedure}

\renewcommand{\thefigure}{S1}
\begin{figure*}[t!]
	\begin{subfigure}[b]{\sfs}
		\includegraphics[width=\linewidth]{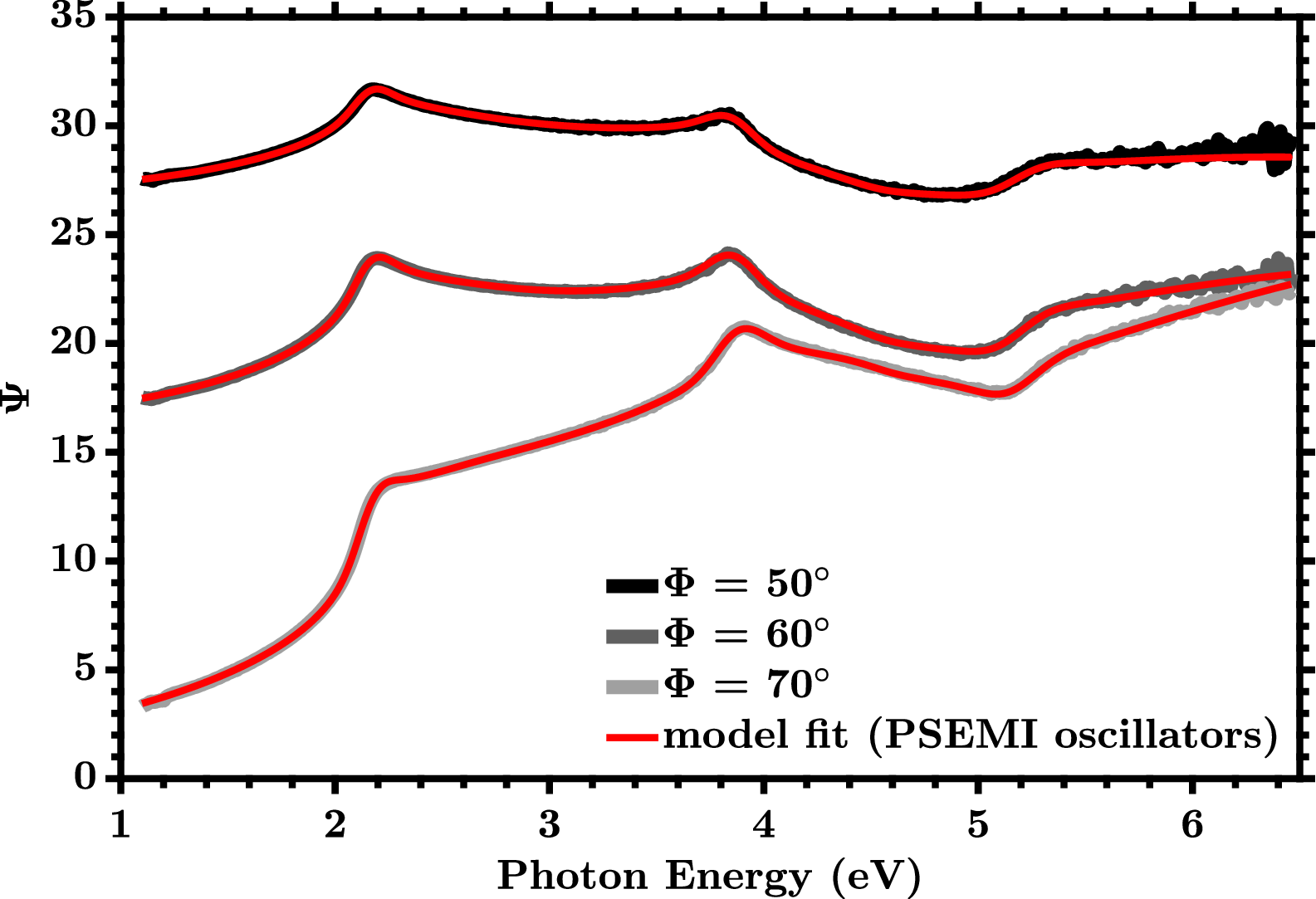}
		\subcaption{}
		\label{UV Psi model}
	\end{subfigure}
\hfil
	\begin{subfigure}[b]{\sfs}
		\includegraphics[width=\linewidth]{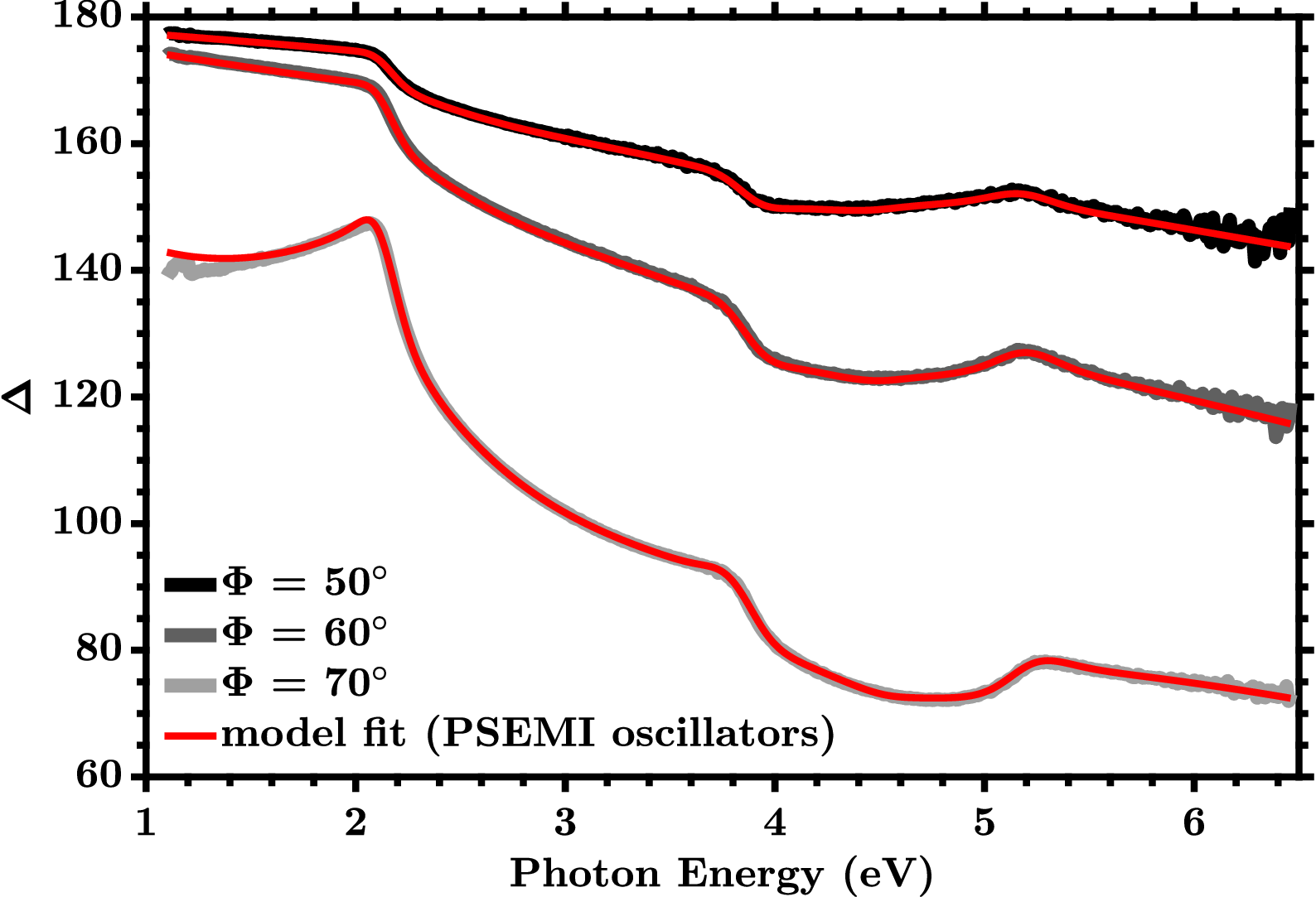}
		\subcaption{}
		\label{UV Delta model}
	\end{subfigure}
	\caption{Parametric semiconductor (PSEMI) oscillator model fits applied to (a) $\Psi$ and (b) $\Delta$ of UVSE before point-by-point fitting.}
	\label{UV models}
\end{figure*}

 In the NIR/VIS/UV spectral range we model two layers: (I) a surface roughness layer, which we model by using a Bruggemann effective-medium layer with a void/layer ratio of 50\% and a variable layer thickness and (II) the ScN film layer, which is constructed by various oscillators as shown in Fig. 3b in the main article. Due to the large thickness of our ScN films we treat them as bulk material and therefore apply no substrate model. In Figs. \ref{UV Psi model} and \ref{UV Delta model} it is obvious, that already the applied model matches the experimental data nearly perfectly and hence, since the WVASE model functions are inherently Kramers-Kronig-consistent, the NIR-UV dielectric function determined from point-by-point fitting is Kramers-Kronig-consistent. In the IR/FIR spectral range only a single layer for the ScN film is used. The bulk approach is further supported by the absence of Fabry-Perot-oscillations in the NIR/VIS range and the absence of sapphire substrate contributions in the MIR range (see Figs. 2a and 2b in the main article). \\

\section*{SIII. Raman spectra line shape fits}
 For quantitative analysis we determine the energy positions by line shape fits.  There, we deal with two problems: (I) the underlying luminescence signal and (II) the non-Lorentzian line shape of some signals. The non-Lorentzian line shape can either arise from coupling with free charge carriers or from even number TO phonons. As discussed in the main article, previous XRD results suggest that those signal possibly stem from the even number TO lines because they exhibit $2\omega_\mathrm{TO}\approx\omega_\mathrm{LO}$. The first order Raman scattering intensity typically exhibits Lorentzian type line shapes, given by
 	\begin{align}
 		I(\omega) = \frac{A\gamma^2}{4(\omega-\omega_0)^2+\gamma^2} \label{lorentz_common}
 	\end{align}
 	with amplitude $A$, eigenfrequency $\omega_0$, and broadening $\gamma$. We apply a local linear background, fitting each peak individually for best fit results. In total, we have 
\begin{align}
	I(\omega) = I_0 + m\omega + \frac{A\gamma^2}{4(\omega-\omega_0)^2+\gamma^2} \label{lorentz_func}
\end{align}
where $m$ is an arbitrary but linear slope and $I_0$ a constant background. \\

In Figs. \ref{473 1TO}-\ref{473 4LO} fit results of all evaluated phonon lines are shown. The applied model matches well with the data, only for the 2LO signal small deviations are visible. \\

\renewcommand{\thefigure}{S2}
	\begin{figure*}
		\begin{subfigure}[b]{\tsfs}
			\centering
			\subcaption{}
			\includegraphics[width=\linewidth]{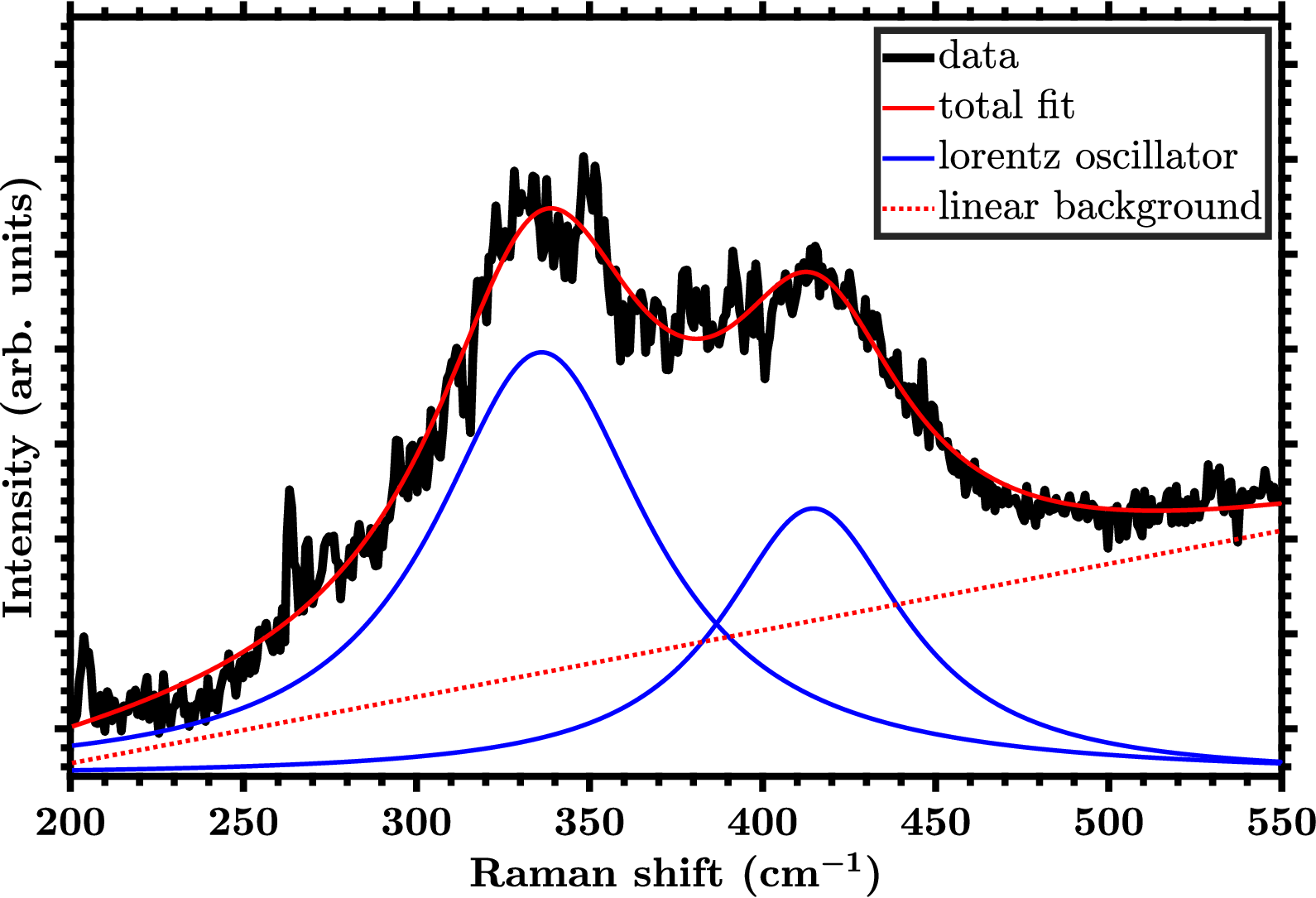} \label{473 1TO}
		\end{subfigure}
	\hfil
	\begin{subfigure}[b]{\tsfs}
		\centering
		\subcaption{}
		\includegraphics[width=\linewidth]{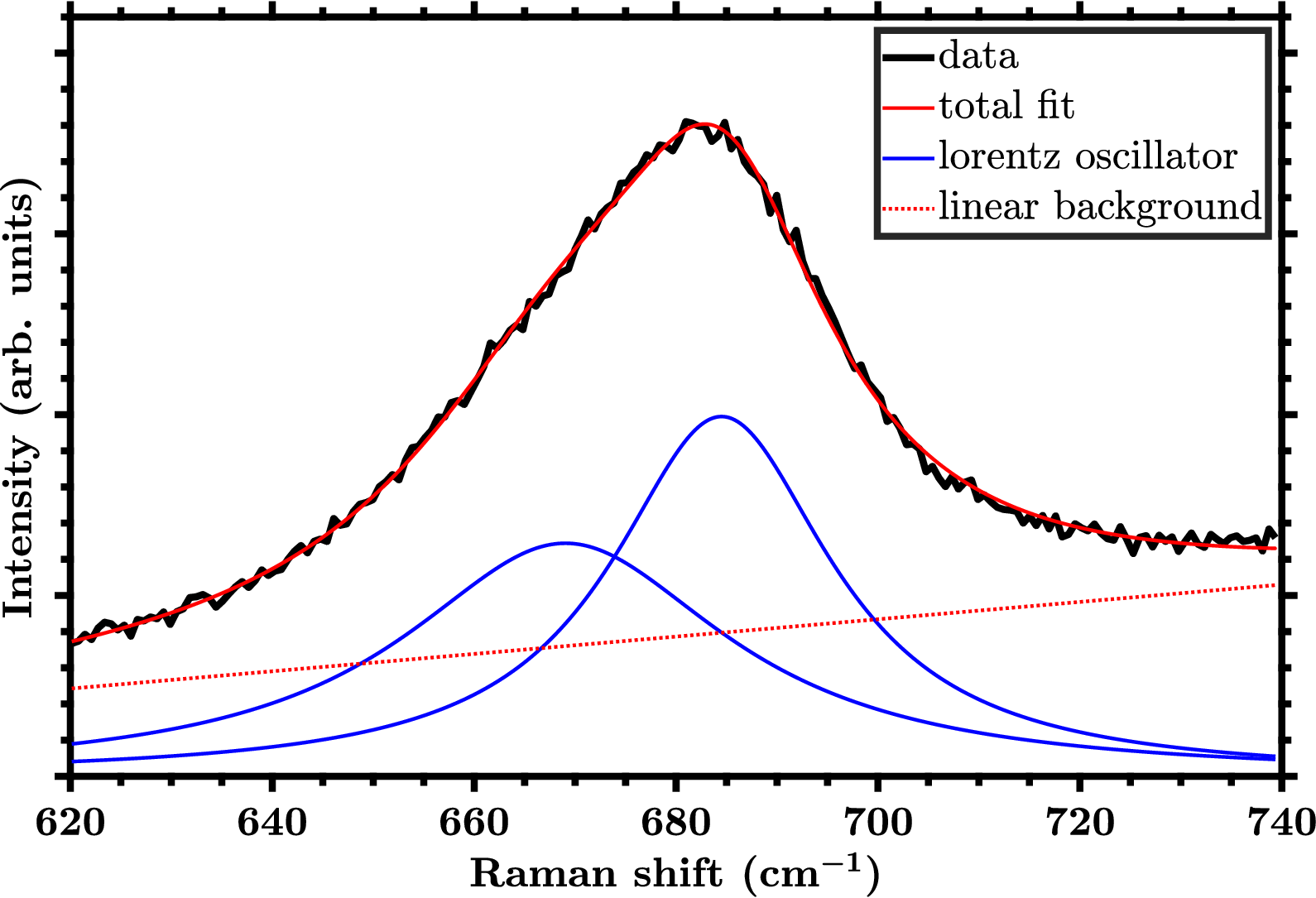} \label{473 1LO}
	\end{subfigure}
	\hfil
	\begin{subfigure}[b]{\tsfs}
		\centering
		\subcaption{}
		\includegraphics[width=\linewidth]{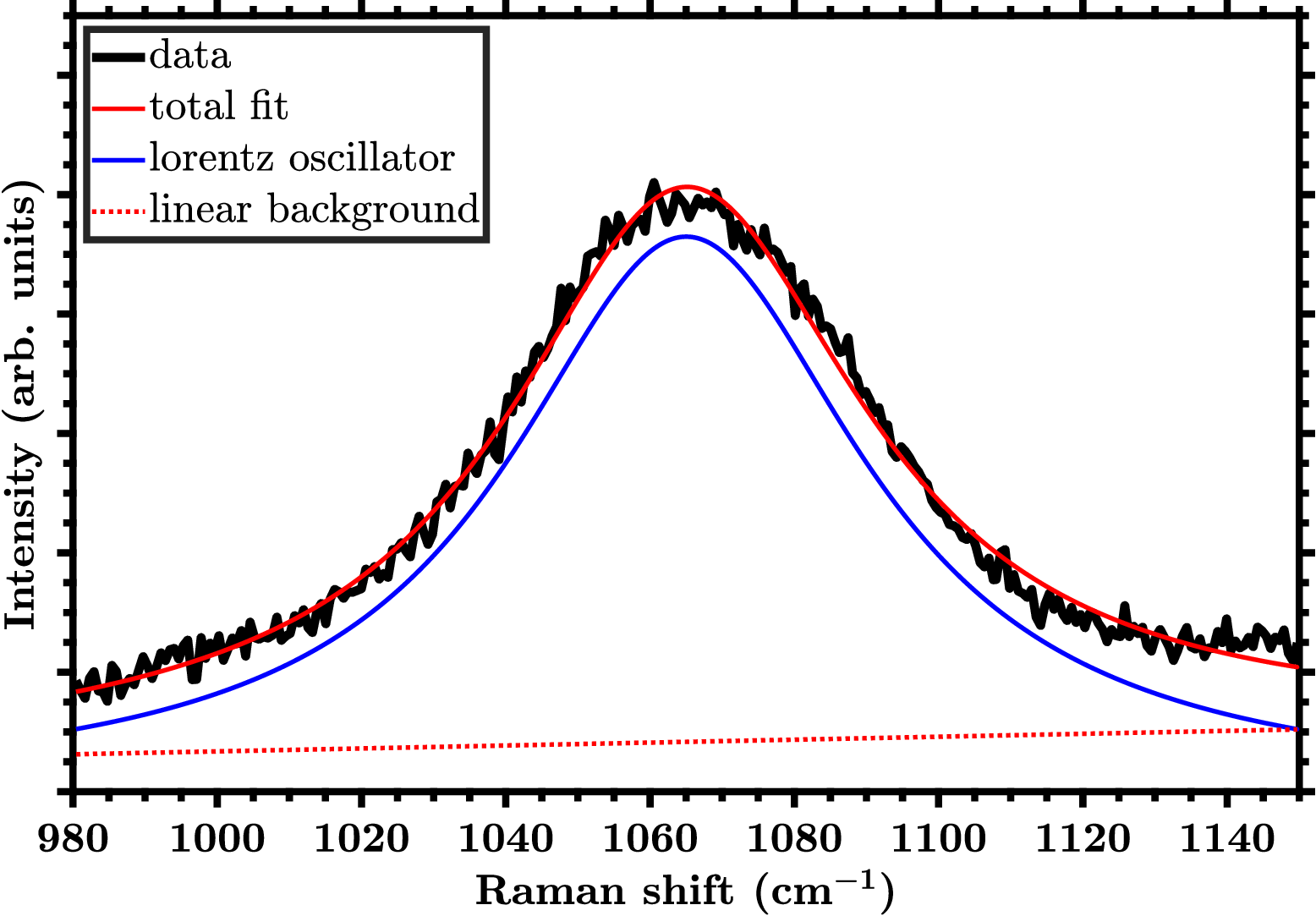}\label{473 3TO}
	\end{subfigure}
	\hfil
	\begin{subfigure}[b]{\tsfs}
		\centering
		\subcaption{}
		\includegraphics[width=\linewidth]{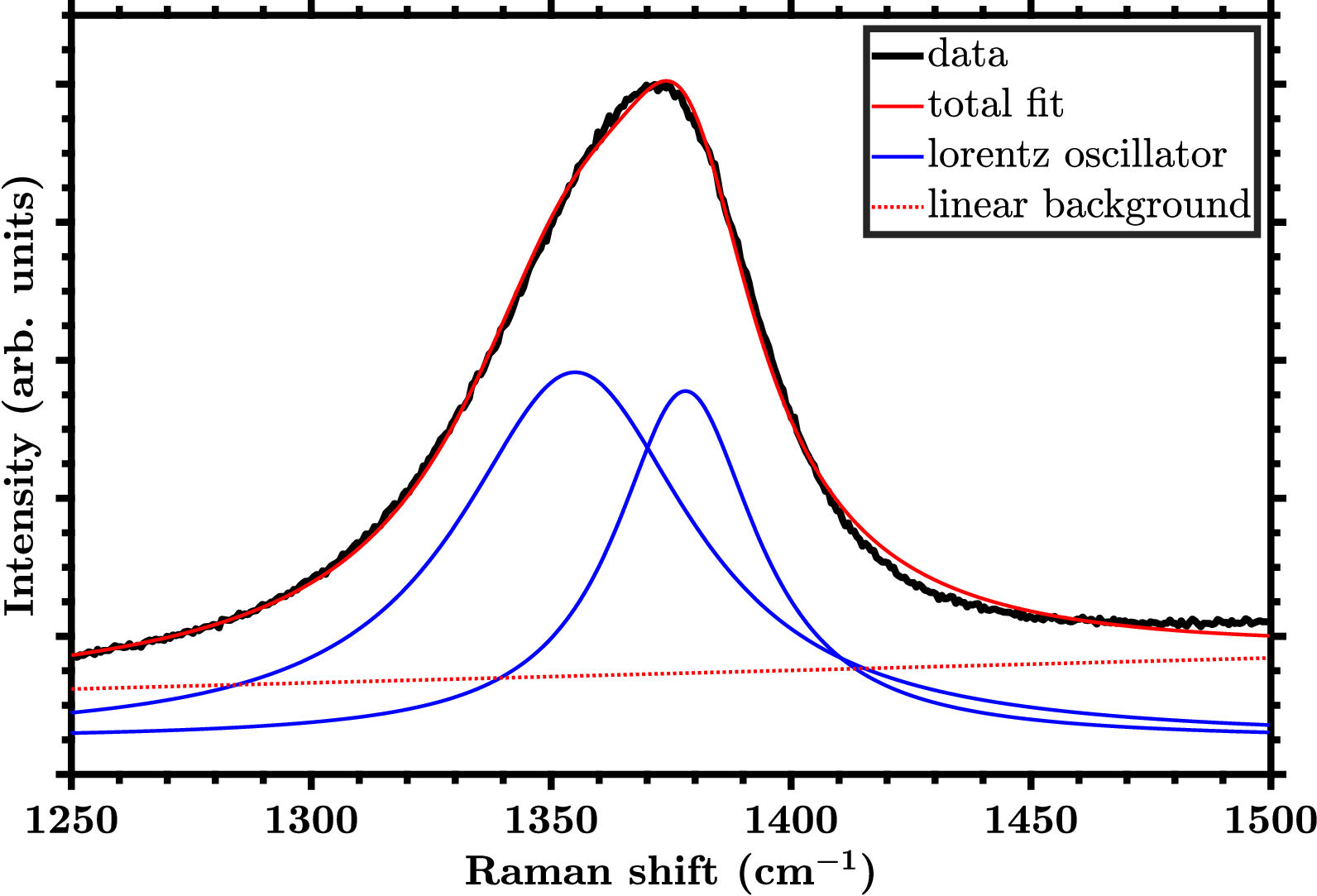}\label{473 2LO}
	\end{subfigure}
	\hfil
	\begin{subfigure}[b]{\tsfs}
		\centering
		\subcaption{}
		\includegraphics[width=\linewidth]{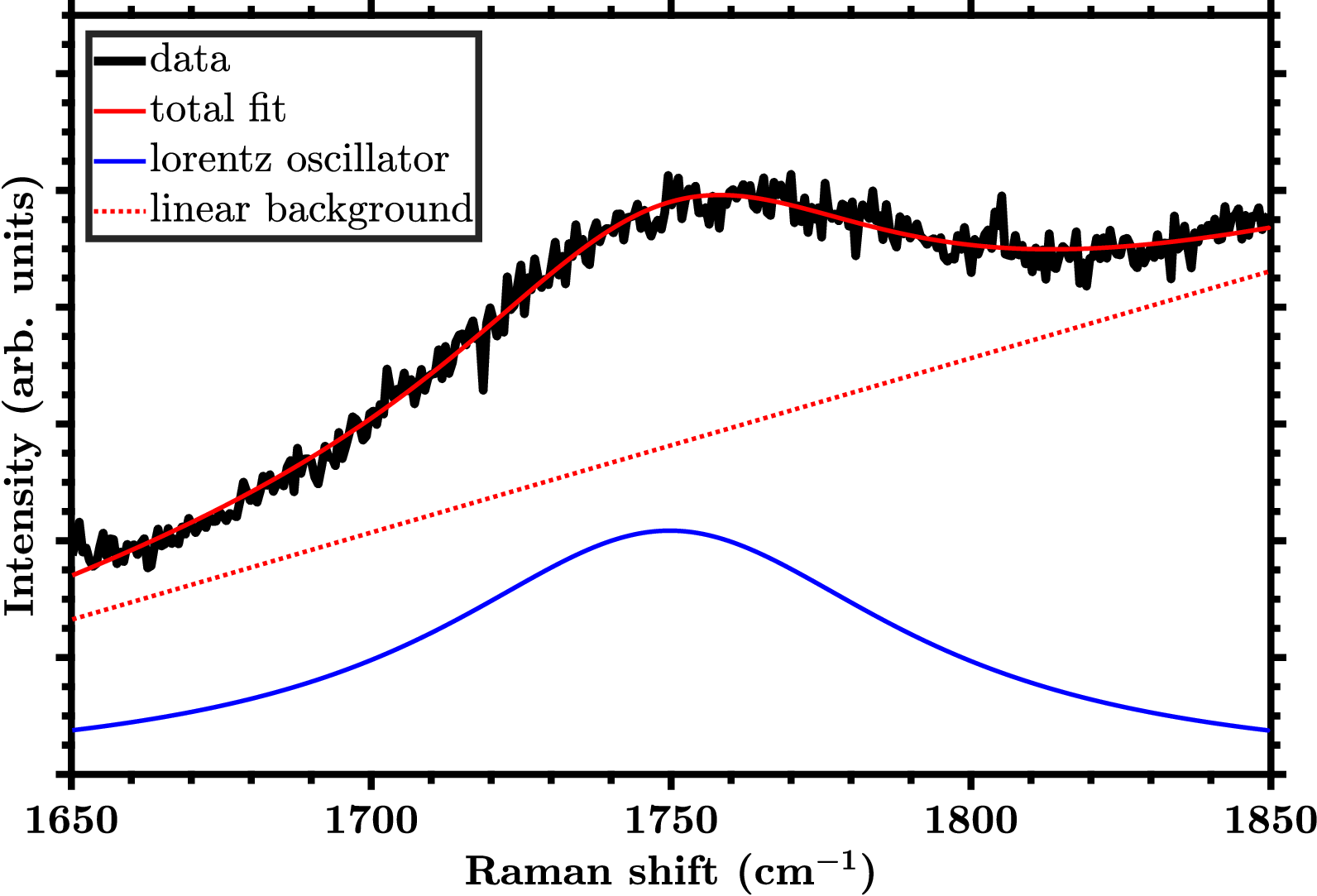}\label{473 5TO}
	\end{subfigure}
	\hfil
	\begin{subfigure}[b]{\tsfs}
		\centering
		\subcaption{}
		\includegraphics[width=\linewidth]{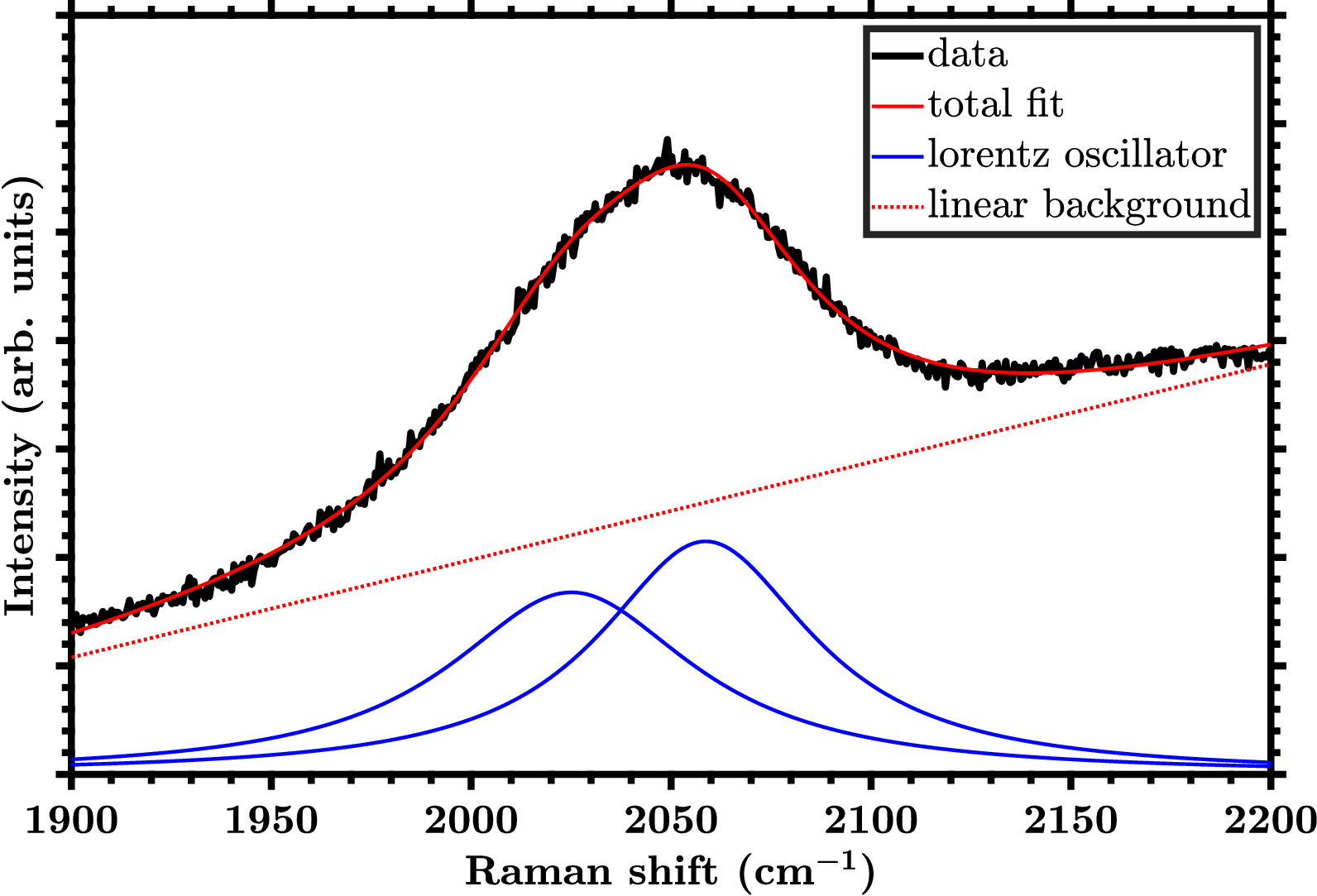}\label{473 3LO}
	\end{subfigure}
	\hfil
	\begin{subfigure}[b]{\tsfs}
		\centering
		\subcaption{}
		\includegraphics[width=\linewidth]{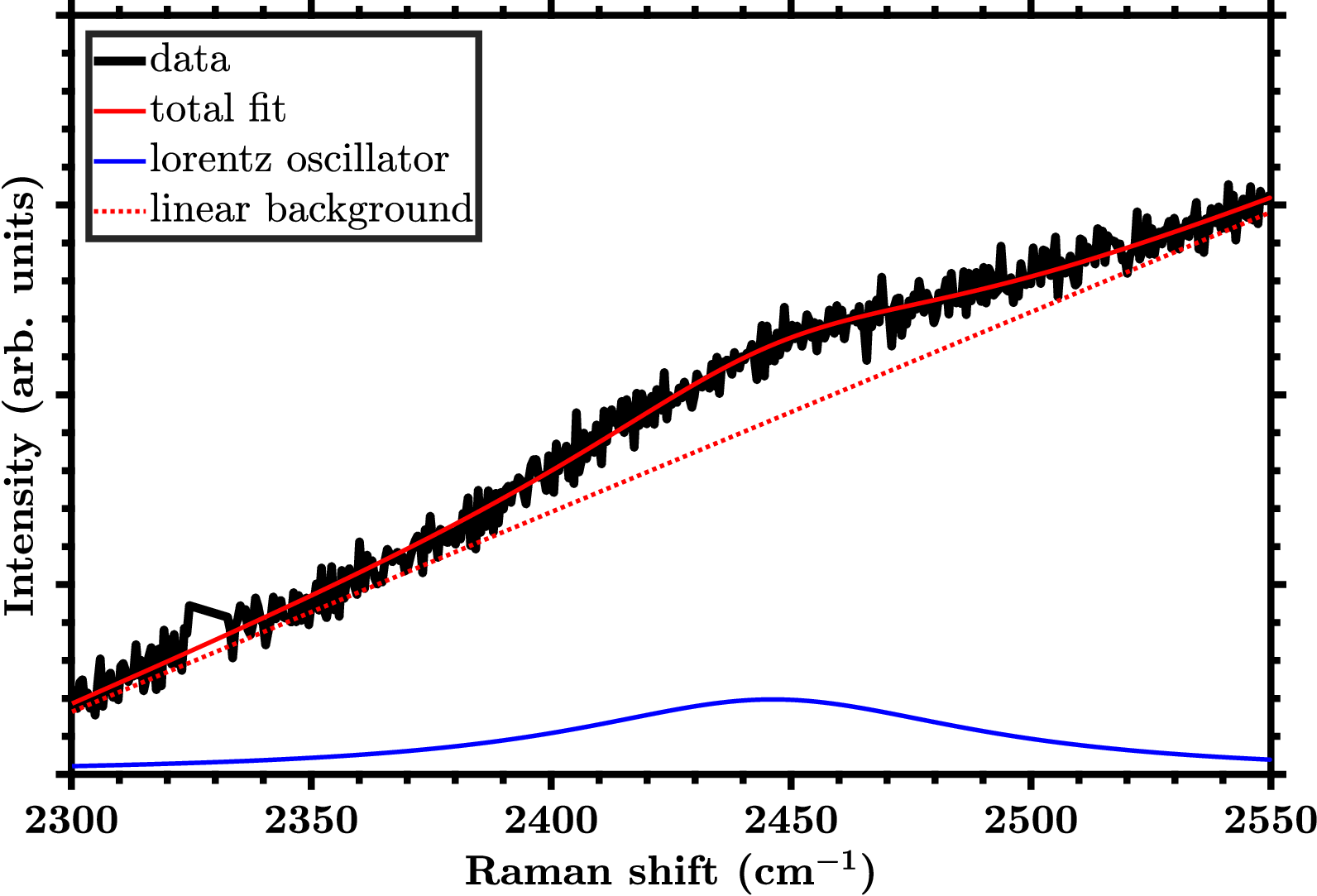}\label{473 7TO}
	\end{subfigure}
	\hfil
	\begin{subfigure}[b]{\tsfs}
		\centering
		\subcaption{}
		\includegraphics[width=\linewidth]{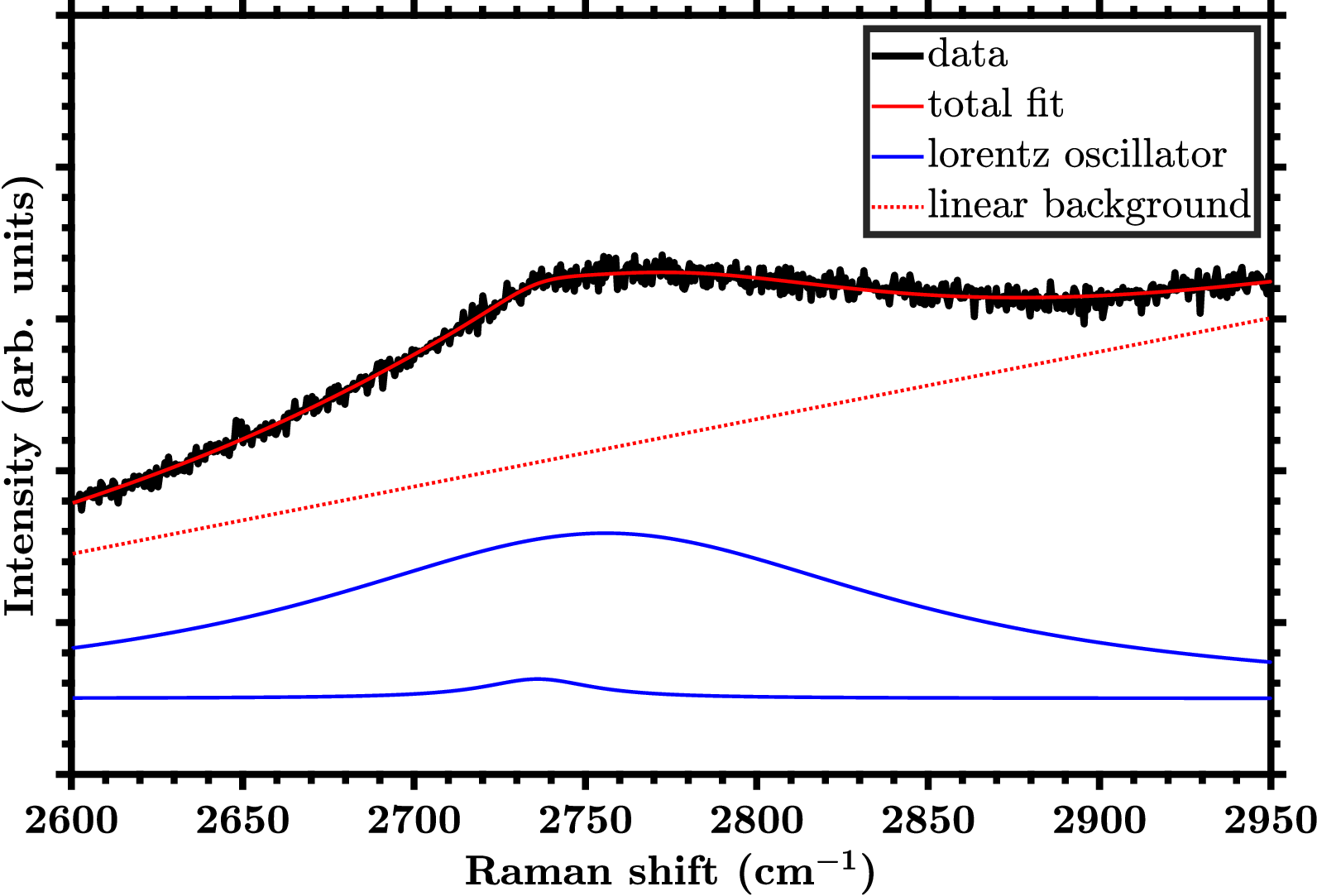}\label{473 4LO}
	\end{subfigure}
\caption{Fit results at $n$LO/TO-spectral regions of the Raman spectrum with $E_{\text{Laser}}=\SI{2.62}{\eV}$.}
\label{Raman 473 fit res}
	\end{figure*}

\section*{SIV. errors}
For values directly extracted as fit parameters the given errors in Tabs. 1 and 2 of the main article display the 99.5\% confidence bounds of the fit. For the derived parameters $\varepsilon_{\infty}$, $\varepsilon_{\text{stat}}$, $Z^*$, $\omega_\text{LO}^{(\text{LST})}$ and the LST-relation probe, the errors are calculated using linear error propagation law and the corresponding fit errors of the required parameters. For the transition energies $E_{\Gamma}$ and $E_{\Gamma'}$, which display via splinefit determined inflection points of $\varepsilon_2$, the error is set as the spectral resolution of the UV ellipsometer at the corresponding photon energy (see Fig. \ref{Resolution UV ellips}). For the indirect bandgap the error is an estimate for our scale-reading precision when deriving it from $\braket{\varepsilon_1}$.

\renewcommand{\thefigure}{S3}
\begin{figure}[H]
	\centering
	\includegraphics[width=\sfs]{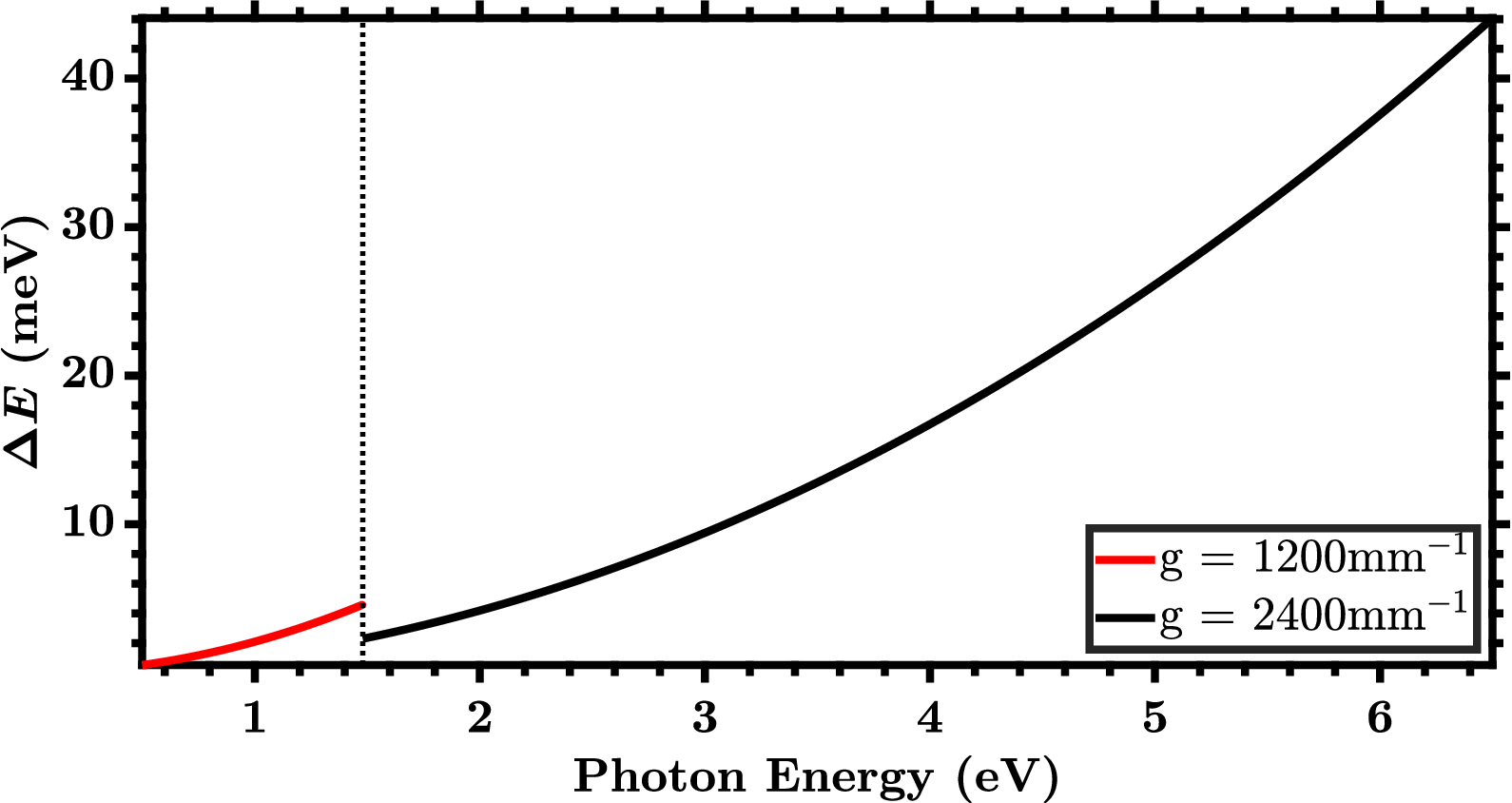}
	\caption{Spectral resolution trace of the UVSE machine. Note, that we use auto slit mode but calculated the error with a constant $d_{\text{slit}}=\SI{500}{\micro\m}$, which is the maximum allowed slit width.}
	\label{Resolution UV ellips}
\end{figure}

%\newpage

	\bibliography{ScN_quasi_bulk_opt_props.bib}